\documentclass[structabstract]{aa}  
\usepackage{graphicx}
\usepackage{xcolor}

\usepackage{amsmath}
\usepackage{txfonts}
\usepackage{natbib}
\bibpunct{(}{)}{;}{a}{}{,} % to follow the A&A style
\begin{document}
   \title{Injection of thermal and suprathermal seed particles into coronal shocks of varying obliquity}
   \titlerunning{Injection of thermal and suprathermal seed particles into coronal shocks of varying obliquity}

   %\authorrunning{M. Battarbee et al}
   \author{M. Battarbee
          \inst{1}
          \and
          R. Vainio\inst{2}
          \and
          T. Laitinen\inst{3}
          \and
          H. Hietala\inst{4}
          }

   \institute{Department of Physics and Astronomy,
              University of Turku, Finland \\
              \email{markus.battarbee@utu.fi}
         \and
             Department of Physics, University of Helsinki, Finland
         \and
             Jeremiah Horrocks Institute, University of Central Lancashire, United Kingdom
         \and
             The Blackett Laboratory, Imperial College, London SW7 2AZ, United Kingdom.
             }

   \date{Accepted manuscript after language edits, 5 September 2013}

\abstract
   {Diffusive shock acceleration in the solar corona can accelerate solar energetic particles to very high energies. Acceleration efficiency is increased by entrapment through self-generated waves, which is highly dependent on the amount of accelerated particles. This, in turn, is determined by the efficiency of particle injection into the acceleration process.}
   {We present an analysis of the injection efficiency at coronal shocks of varying obliquity. We assessed injection through reflection and downstream scattering, including the effect of a cross-shock potential. Both quasi-thermal and suprathermal seed populations were analysed. We present results on the effect of cross-field diffusion downstream of the shock on the injection efficiency.}
   {Using analytical methods, we present applicable injection speed thresholds that were compared with both semi-analytical flux integration and Monte Carlo simulations, which do not resort to binary thresholds. Shock-normal angle $\theta_{B\mathrm{n}}$ and shock-normal velocity $V_\mathrm{s}$ were varied to assess the injection efficiency with respect to these parameters.}
   {We present evidence of a significant bias of thermal seed particle injection at small shock-normal angles. We show that downstream isotropisation methods affect the $\theta_{B\mathrm{n}}$-dependence of this result. We show a non-negligible effect caused by the cross-shock potential, and that the effect of downstream cross-field diffusion is highly dependent on boundary definitions.}
   {Our results show that for Monte Carlo simulations of coronal shock acceleration a full distribution function assessment with downstream isotropisation through scatterings is necessary to realistically model particle injection. Based on our results, seed particle injection at quasi-parallel coronal shocks can result in significant acceleration efficiency, especially when combined with varying field-line geometry.}


   \maketitle
%
%________________________________________________________________

\section{Introduction}

The majority of energetic particles in the near-Earth space environment are generated by energetic processes at the Sun \citep{1999SSRv...90..413R}. Coronal or interplanetary shocks driven by coronal mass ejections (CMEs) play a major role in accelerating seed particles to high energies through a process of repeated shock encounters. This process of diffusive shock acceleration (DSA) was initially described by \cite{1977ICRC...11..132A}, \cite{1977DoSSR.234R1306K}, \cite{1978ApJ...221L..29B}, and \cite{1978MNRAS.182..147B}. Repeated shock encounters are assumed to occur due to trapping by upstream Alfv\'enic turbulence generated by the accelerated particles themselves \citep{1983JGR....88.6109L}, and scattering of particles in the turbulent downstream medium.

For ambient solar wind particles to take part in the DSA process, they have to return to the upstream after encountering the propagating coronal shock for the first time. This process of \emph{injection} is straightforward to understand assuming the seed particles have kinetic energies well in excess of the mean upstream incident ram energy. However, quantitative understanding of injection into DSA at low energies remains a challenge. A comprehensive understanding of the injection efficiency is especially important when particle entrapment in DSA is attributed to self-generated waves (see, e.g., \citealp{1994ApJ...424.1032N}, \citealp{2003A&A...406..735V}, \citealp{2005ApJS..158...38L}, \citealp{2008ApJ...686L.123N} and \citealp{2007ApJ...658..622V,2008JASTP..70..467V}). This is because self-generated wave intensity is scaled by the absolute level of resonant particle intensity. Thus, from the particle point of view, the theory is non-linear.

In this paper, we investigate seed particle injection using a simplified shock model, suitable for implementation in self-consistent Monte Carlo simulations of DSA (see, e.g., \citealp{2011A&A...535A..34B}). We present injection probabilities for quasi-thermal and non-thermal particle populations. We study only the injection of particles to the upstream after their first interaction with the shock front, not the subsequent energising process through consecutive crossings.

The bulk of the solar wind consists of thermal, cool particles. Much research has been done to link shock acceleration with non-thermal seed populations, which could be remnants from previous energetic eruptions (see, e.g., \citealp{1999ApJ...525L.133M,2001ApJ...558L..59T,2005ApJ...625..474T,2006ApJ...646.1319T,2007ApJ...662L.127S}). It is uncertain whether energetic remnant seed populations are capable of supplying the required particle flux to generate efficient turbulent trapping in front of the shock, especially considering the high energies attained (\citealp{1984JGR....89.2122V}). Given the speeds at which coronal shocks propagate, only a small fraction of thermal ions are likely to be injected into the acceleration process.

In this paper, we study the injection of thermal and suprathermal particles into the shock acceleration process. The quasi-thermal and non-thermal populations are described using $\kappa$-distributions. We used a simplified one-dimensional model to analyse the reflection and transmission of incident particles at the shock front. Comparisons of work presented by \cite{1989MNRAS.239..995K} and \cite{1991MNRAS.249..551O} suggest that the one-dimensional adiabatic approximation is sufficient for our model. Particle transmission uses the scatter-free approximation (see, e.g., \citealp{1988SSRv...48..195D}). However, we propagated the particles in the downstream until they were isotropised through pitch-angle scattering before applying analytical return-probability estimates.

The angle $\theta_{B\mathrm{n}}$ between the shock normal and the upstream magnetic field, with values $\theta_{B\mathrm{n}}=0^\circ \ldots 90^\circ$, was varied to assess the coupling of shock-normal angle $\theta_{B\mathrm{n}}$ and particle injection probability. The electric cross-shock potential was assessed along with its effect on injection and reflection of incident particles. It should be noted that our analysis considers only the adiabatic approximation of shock-drift acceleration and does not take into account the effects of shock surfing.

Modification of shock-velocity profiles due to particle acceleration, as presented for instance in \cite{1996ApJ...473.1029E}, is not expected to play a major role in interplanetary shocks with low Alfv\'enic Mach numbers (see, e.g., \citealp{2006AdSpR..37.1408T}). Therefore we adopted the test-particle approximation.

At supercritical shocks (see, e.g., \citealp{2011ApJ...739L..64B}), specularly reflected ions, ion beams, and upstream waves can interact and cause shock reformation \citep{1984AdSpR...4..231L,1989GeoRL..16..345B,1992JGR....97.8319S} and can also affect the energy dissipation. Because specular reflection requires full-orbit simulations, we did not consider the effect of shock reformation on particle injection.

%__________________________________________________________________

\section{Particle injection}

In previous studies of shock acceleration, particle injection has been modelled through simplifications such as mono-energetic (e.g., \citealp{2012AdSpR..49.1067L}) or power-law (e.g., \citealp{2007ApJ...662L.127S,2008ApJ...686L.123N}) seed populations, or by using a thermostat model \citep{1995A&A...300..605M,1998AdSpR..21..551M}. \cite{2012ApJ...757...97N} employed a $\kappa$-distribution, using an ad-hoc minimum injection energy. A classical approximation for minimum particle injection speed is $v > u_1$ (see, e.g., \citealp{1983JGR....88.6109L} and \citealp{2009ApJ...693..534L}), where $v$ is the particle speed and $u_1$ is the plasma flow speed along the upstream magnetic field line. We show that this threshold speed is not always a good approximation. 
 
Particle injection at oblique shocks has been simulated through Monte Carlo methods for instance by \cite{1991MNRAS.249..551O}, \cite{1993ApJ...409..327B}, and \cite{1995ApJ...453..873E}. In the context of DSA at coronal shocks, we limit our analysis to a single shock encounter instead of propagating particles, for example, until they reach a threshold energy (see, e.g., \citealp{1994ApJS...90..547B}).

Highly variable magnetic fields and their effect on particle injection have been studied among others by \cite{2005ApJ...624..765G} and \cite{2006JPhCS..47..160G}. At large shock-normal angles ($\theta_{B\mathrm{n}} \gtrsim 85^\circ$), such as at the quasi-perpendicular termination shock, cross-field diffusion via field-line random walk across the shock plays a major role, and the approximations presented here may not be applicable.

%__________________________________________________________________

\section{Shock geometry} \label{sec:shockgeometry}

Shocks launched from the solar corona travel outwards in the heliosphere, where the background magnetic field is described as the Parker spiral. A propagating shock can intersect magnetic field lines with a multitude of shock-normal angles $\theta_{B\mathrm{n}}$, especially during expansion within the complex coronal field. The specifics of shock expansion in the coronal magnetic field have been detailed for instance by \cite{2006A&A...455..685S}, \cite{2009A&A...507L..21S}, and \cite{2011ASTRA...7..387P}. In this paper, we fixed the shock-normal propagation velocity $V_\mathrm{s}$ (measured in the inertial frame) and the solar wind parameters. We varied the angle $\theta_{B\mathrm{n}}$ between the propagation direction of the shock and the mean magnetic field to assess injection efficiency at different points of a curved shock front.

We considered different shock-normal angles $\theta_{B\mathrm{n}}$ as distinct test cases, ranging from $\theta_{B\mathrm{n}}=0^\circ$ to $\theta_{B\mathrm{n}}=60^\circ$. This means that our results only apply to quasi-parallel and oblique portions of coronal or interplanetary shocks, but not to quasi-perpendicular shocks. We also neglected changes to the upstream shock-normal angle caused for example by upstream wave growth (see, e.g., \citealp{1993JGR....98...47S}) or field-line meandering, and limited our cases to constant values of $\theta_{B\mathrm{n}}$. However, our results for each shock-normal angle can be applied as short intervals of locally constant $\theta_{B\mathrm{n}}$, as part of an evolving, propagating shock front.

We examined plasma flows at the shock in the de Hoffmann--Teller frame, where the mean magnetic field and mean plasma flow direction are parallel to each other in both the upstream and the downstream of the shock. Using the Rankine--Hugoniot equations, we solved the gas and magnetic compression ratios $r_\mathrm{g}$ and $r_{B}$ at the shock front.

In our notation, de Hoffmann -- Teller frame upstream quantities are sub-scripted with a $1$ and downstream quantities with a $2$. Positive pitch-angle cosines $\mu$ indicate propagation towards the upstream, whereas negative values indicate propagation towards the downstream. Quantities given in the upstream plasma frame are bare, and in the downstream plasma frame primed. The downstream plasma frame quantities are of most use when assessing return probabilities, and only upstream plasma frame quantities are initially given. The solar wind flows radially outward with velocity $u_\mathrm{sw}$ along the radial magnetic field. The plasma flow velocity $u_1$ is given parallel to the magnetic field, with $u_1 = V_\mathrm{s} \sec \theta_{B\mathrm{n}} -u_\mathrm{sw}$.

Our radial solar wind model, introduced in \cite{2010AIPC.1216...84B}, satisfies required continuities of mass flux and magnetic flux. Our choice of heliocentric distance $r=3.17 ~R_\odot$ results in a high Alfv\'en speed $v_\mathrm{A}$. Upstream parameters include the plasma density $\rho = 3.48\cdot10^5 \,\mathrm{cm}^{-3}$, background magnetic field flux density $B_1 \approx 0.19\,\mathrm{G}$, temperature $T \approx 2.0 \cdot 10^{6} \,\mathrm{K}$, sound speed $c_\mathrm{s} \approx 2.34\cdot 10^{7} \,\mathrm{cm\,s}^{-1}$, Alfv\'en speed $v_\mathrm{A} \approx 6.97\cdot 10^{7}\,\mathrm{cm\,s}^{-1}$, the plasma beta $\beta_1=2 \gamma_\mathrm{p}^{-1}c_\mathrm{s}^{2}v_\mathrm{A}^{-2}=0.135$, solar wind speed $u_\mathrm{sw} \approx 9.98 \cdot 10^{6}\,\mathrm{cm\,s}^{-1}$, and the wave group speed $V=u_\mathrm{sw}+v_\mathrm{A}\approx 7.97 \cdot 10^{7}\,\mathrm{cm\,s}^{-1}$. Here $\gamma_\mathrm{p}=5/3$ is the ratio of specific heats. For calculations and simulations, we used two different shock-normal velocities of $V_\mathrm{s} = 1500\,\mathrm{km\,s}^{-1}$ and $V_\mathrm{s} = 2000\,\mathrm{km\,s}^{-1}$ with shock-normal angles ranging from $\theta_{B\mathrm{n}} = 0^\circ$ to $\theta_{B\mathrm{n}} = 60^\circ$. 

In table \ref{table:speeds}, we present values for the upstream flow speed $u_{1}$, the Alfv\'enic Mach number $M_\mathrm{A}$, the gas compression ratio $r_\mathrm{g}$, the magnetic compression ratio $r_{B}$, and the downstream shock-normal angle $\theta_{B\mathrm{n},2}$ for the cases studied in this paper. The shock compression ratio was solved parametrically as presented in \cite{1999A&A...343..303V}, with the case $\theta_{B\mathrm{n}}=0$ solved for upstream angle $0.03^\circ$, to make certain that the shock is a fast-mode shock.

\begin{table}[htb]
\caption{Shock-normal velocity, angle, and shock parameters solved with Rankine-Hugoniot equations. Speeds $V_\mathrm{s}$ and $u_1$ are given in units km s$^{-1}$ and angles are given in degrees. Note that the parallel-shock case is computed for $\theta_{B\mathrm{n}} = 0.03^\circ$.}
\vspace{4mm}
\centering
\begin{tabular}{c}
$V_\mathrm{s} = 1500\,km\,s^{-1}$
\end{tabular} \\
\begin{tabular}{ c c c c c c }
\hline
$\theta_{B\mathrm{n}}$ & $u_{1}$ & $M_\mathrm{A}$ & $r_\mathrm{g}$ & $r_{B}$ & $\theta_{B\mathrm{n},2}$ \\ \hline
0    & 1400 & 2.00 & 3.68 & 1.01 & 5.8 \\ \hline
2.5  & 1400 & 2.00 & 3.43 & 1.26 & 37.5 \\ \hline
5    & 1410 & 2.01 & 3.22 & 1.44 & 46.1 \\ \hline
7.5  & 1413 & 2.02 & 3.06 & 1.56 & 50.4 \\ \hline
10   & 1420 & 2.04 & 2.93 & 1.65 & 53.2 \\ \hline
12.5 & 1436 & 2.06 & 2.83 & 1.71 & 55.3 \\ \hline
15   & 1450 & 2.08 & 2.74 & 1.77 & 56.9 \\ \hline
20   & 1496 & 2.14 & 2.59 & 1.85 & 59.5 \\ \hline
25   & 1555 & 2.22 & 2.48 & 1.91 & 61.7 \\ \hline
30   & 1630 & 2.33 & 2.40 & 1.95 & 63.7 \\ \hline
45   & 2020 & 2.89 & 2.24 & 2.03 & 69.7 \\ \hline
60   & 2900 & 4.14 & 2.17 & 2.09 & 76.1 \\ \hline
\end{tabular} \\ 
\vskip4mm
\begin{tabular}{c}
$V_\mathrm{s} = 2000\,km\,s^{-1}$
\end{tabular} \\
\begin{tabular}{ c c c c c c }
\hline
$\theta_{B\mathrm{n}}$ & $u_{1}$ & $M_\mathrm{A}$ & $r_\mathrm{g}$ & $r_{B}$ & $\theta_{B\mathrm{n},2}$ \\ \hline
0    & 1900 & 2.72 & 3.83 & 1.00 & 1.2 \\ \hline
2.5  & 1900 & 2.72 & 3.81 & 1.04 & 16.5 \\ \hline
5    & 1910 & 2.73 & 3.76 & 1.15 & 29.9 \\ \hline
7.5  & 1920 & 2.74 & 3.69 & 1.29 & 39.6 \\ \hline
10   & 1930 & 2.76 & 3.62 & 1.43 & 46.5 \\ \hline
12.5 & 1949 & 2.79 & 3.53 & 1.56 & 51.4 \\ \hline
15   & 1970 & 2.82 & 3.45 & 1.69 & 55.0 \\ \hline
20   & 2028 & 2.90 & 3.30 & 1.89 & 60.2 \\ \hline
25   & 2107 & 3.01 & 3.17 & 2.05 & 63.7 \\ \hline
30   & 2210 & 3.16 & 3.07 & 2.17 & 66.5 \\ \hline
45   & 2730 & 3.90 & 2.84 & 2.41 & 72.9 \\ \hline
60   & 3900 & 5.58 & 2.71 & 2.54 & 78.6 \\ \hline
\end{tabular}
\vskip4mm
\label{table:speeds}
\end{table}

\subsection{Cross-shock potential} \label{sec:cross-shock}

The solar wind consists of ions and electrons in suitable quantities to achieve macroscopic plasma neutrality. An ion-to-electron imbalance at kinetic scales forms an electrostatic field across the shock profile, resulting in the cross-shock potential. As presented for instance in \citet{2000JGR...10520957H}, this cross-shock potential can be approximated by various methods. Although we focused on coronal shocks, we used the estimates found from observing the bow shock of the Earth, where the potential jump is set to
\begin{align}
\Delta \Phi = \phi \frac{1}{2}m_\mathrm{p} \left( u_{1,\mathrm{n}}^{2} - u_{2,\mathrm{n}}^{2} \right),
\end{align}
which is proportional to the change of the normal component of proton ram energy. Here $\phi=0.12$ is an observational parameter \citep{2000JGR...10520957H} and $m_\mathrm{p}$ is the proton mass.

Although the significance of this electric potential has been investigated previously (see, e.g., \citealp{1996JGR...101.4871G}, \citealp{2001PhPl....8.4560Z} and \citealp{2013ApJ...767....6Z}), its effect on particle acceleration efficiency is still largely ignored. Unless otherwise stated, the cross-shock potential is always accounted for in our calculations.

\section{Seed particle populations} \label{sec:seedpopulations}

\begin{figure}[!t]
\centering
\resizebox{\hsize}{!}{\includegraphics{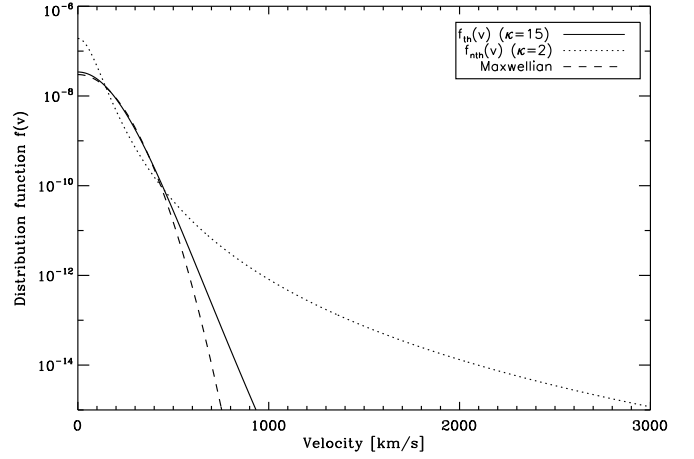}}%pics/kappa_v_distribution
\caption{Particle distribution functions $f_\mathrm{th}(v)$, $f_\mathrm{nth}(v)$ and a Maxwellian distribution at $T \approx 2.0 \cdot 10^{6} \,\mathrm{K}$.}
%$f_\mathrm{th}(v)+f_\mathrm{nth}(v)$
\label{fig:kappadist}
\end{figure}

The bulk of the ambient solar wind consists of protons and electrons. Although most particles encountered by a coronal shock have left the surface of the Sun in a thermalised state (as depicted by a Maxwellian velocity distribution), the ambient plasma contains an additional suprathermal halo. As the corona exists in a collisionless state, this halo remains unthermalised during expansion to interplanetary scales. We describe particle distributions using a $\kappa$-distribution \citep{1968JGR....73.2839V}, where $\kappa$ is a model parameter defining the strength of the suprathermal tail. The representative speed parameter
\begin{equation}
  w_{0}=\sqrt{2 T k_\mathrm{B} \frac{\kappa-1.5}{\kappa m_\mathrm{p}} } \label{eq:speedparameter}
\end{equation}
is based on the ion temperature $T$, where $k_\mathrm{B}$ is the Boltzmann constant. The particle distribution function is
\begin{equation}
  f(v) = \frac{ n(r) \Gamma(\kappa+1) }{ w_{0}^{3}\pi^{3/2}\kappa^{3/2}\Gamma(\kappa-1/2) }
  \left[1+ \frac{v^2}{\kappa w_{0}^{2}}\right]^{-\kappa-1} \label{eq:kappa},
\end{equation}
where $\Gamma(x)$ denotes the gamma function. We compared the injection efficiency for a distribution with a strong suprathermal tail ($\kappa=2$) and for a near-Maxwellian distribution ($\kappa=15$). These populations along with a Maxwellian distribution are shown in figure \ref{fig:kappadist}.

\subsection{Equation of motion} \label{sec:eqnofmotion}

The equation of motion for a system of particles undergoing elastic pitch-angle scattering in the frame of flowing plasma can be simplified, considering motion within a flux tube extending from the shock $(x_1=0)$ to the far upstream $(x_1\rightarrow\infty)$. The plasma speed $u_1$ is assumed constant and aligned with the constant upstream magnetic field. In a steady state, the particle distribution function obeys
\begin{equation}
(\mu v-u_{1})\frac{\partial f}{\partial x_1}=\frac{\partial}{\partial\mu}D_{\mu\mu}\frac{\partial f}{\partial\mu},
\end{equation}
where $x_1$ is the distance measured along the field lines in the shock frame. $f$ is taken to be a function of mixed coordinates in the frame of the scattering centres ($v,\mu$) and the shock ($x_1$).

\subsection{Flux conservation} \label{sec:fluxconservation}

When modelling the injection of strongly suprathermal particles at a coronal shock, information of the approaching shock can propagate into the upstream and alter the incident particle population. To maintain the flux conservation, the incident particle pitch-angle distribution should be adjusted. A full description of the steps taken is included in Appendix \ref{appsec:fluxweighting}.

According to Liouville's theorem, the shock-incident particle phase-space distribution $f({\bf x},{\bf p},t)$ is conserved along particle orbits. Conservation of mass implies that the shock-normal integrated particle flux
\begin{align*}
F = \cos \theta_{B\mathrm{n}} \int (\mu v-u) f({\bf x},{\bf p},t) \, d^3p
\end{align*}
is conserved.

\subsection{Downstream population isotropy} \label{sec:isotropy}

Particles transmitted to the downstream region of the shock scatter off turbulence and isotropise. A portion of the isotropised particles, if energetic enough, can return to the shock and be injected into the upstream. \cite{1991SSRv...58..259J} gave the return probability for an isotropised particle as
\begin{equation}
P_\mathrm{ret} = \begin{cases}
\left(\frac{v'-u_2}{v'+u_2}\right)^2, \quad v'>u_2  \\
0,\hfill v'\leq u_2.
\end{cases}\label{eq:returnprob}
\end{equation}
This probability is the ratio of particle fluxes towards and away from the shock at a plane in the downstream, assuming isotropic particle distributions and a downstream of infinite extent. We applied this equation to the guiding centre motion of a particle along the magnetic field and, thus, the speed $u_2$ in equation \ref{eq:returnprob} is measured in the de Hoffmann -- Teller frame.

The method for estimating the downstream scattering of particles can have a significant effect on the injection probability. Assessing downstream isotropisation methods and their connection to injection efficiency are the primary focus of this paper. One method is deducing the temperature of the transmitted population and choosing injection energies based on the downstream gas temperature (see, e.g., \citealp{1995A&A...300..605M} and \citealp{2007SSRv..130..255Z}). Another is to assume that particles retain their plasma frame speed and achieve isotropy through elastic small- or large-angle scattering in the plasma frame \citep{1991SSRv...58..259J}. This can be either Monte Carlo-simulated or assumed to take place instantaneously. In section \ref{sec:montecarlo}, we present Monte Carlo simulations in which we varied the simulation parameters to find the difference between assuming instant isotropy and letting particles scatter until they have travelled a pre-chosen distance. For simplicity, we employed an isotropic small-angle scattering process with pitch-angle diffusion coefficient
\begin{equation}
D_{\mu\mu}=\frac{v}{2\lambda_\parallel}(1-\mu^{2}),
\end{equation}
where $\lambda_\parallel$ is the mean free path of particles undergoing scattering. The field-parallel diffusion coefficient is $\kappa_\parallel = v'\lambda_\parallel /3$.

\subsection{Cross-field diffusion} \label{sec:crossfield}

Owing to the presumably strong scattering in the downstream, particles are no longer locked to a single magnetic field line (see, e.g., \citealp{1993ICRC....2..243J} and \citealp{1995AdSpR..15..397B}). A strongly turbulent downstream field could be described by field-line random walk, which presents challenges related to a field line potentially intersecting the shock multiple times. A simplification suitable for DSA simulations is to describe the downstream cross-field propagation of a particle as perpendicular diffusion across the mean magnetic field. We describe downstream cross-field diffusion with a pitch-angle-dependent perpendicular diffusion coefficient 
\begin{equation}
\kappa_\perp(\mu') = a_\perp \kappa_\parallel (1-\mu'^2),
\end{equation}
where $a_\perp$ is a cross-field diffusion strength parameter and the term $(1-\mu'^2)$ scales motion according to the Larmor radius $r'_\mathrm{L}=mv'_\perp/(qB')$.

\section{Analytical injection thresholds}\label{sec:analytical}

In this section, we present analytical solutions of particle-speed thresholds, which allow either reflection at a shock front or injection following transmission to the downstream. We transmitted particles across the infinitesimally thin shock using the adiabatic guiding centre approximation and scatter-free approximation, similar to \cite{1988SSRv...48..195D}. However, we extended the formulation to include the cross-shock electric potential $\Delta\Phi$, solving the particle downstream plasma frame speed $v'$ using only shock- and upstream-flow parameters.

\subsection{Particle reflection by the shock structure} \label{sec:reflection}

\begin{figure}[!t]
\centering
\resizebox{\hsize}{!}{\includegraphics{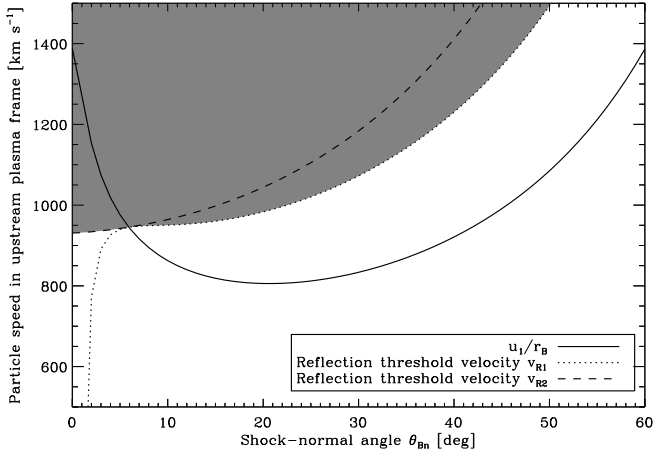}}%pics/speedthresholds_vs1500
\caption{Reflection threshold speeds for $V_\mathrm{s}=1500\,\mathrm{km\,s}^{-1}$: Within the shaded area, protons may be reflected at the shock. Threshold \#1 ($v > v_{\mathrm{R}1}$, $v > u_{1}/r_{B}$) is defined with equation \eqref{eq:magnref1} and threshold \#2 ($v > v_{\mathrm{R}2}$, $v < u_{1}/r_{B}$) with equation \eqref{eq:magnref2}.} 
\label{fig:thresholds}
\end{figure}

The equations for adiabatic scatter-free particle transmission can be used to assess changes in particle motion because it is transmitted across the shock front. A particle is reflected at the shock front if the decrease of its parallel kinetic energy generated by magnetic compression and the cross-shock potential exceeds its initial upstream parallel kinetic energy. Assuming the conservation of magnetic moment and energy in the shock-crossing of a particle, the downstream (shock frame) parallel velocity can be derived as
\begin{equation}
\mu_2 v_2 = -\sqrt{(\mu v-u_{1})^{2}-\tfrac{2}{m}\mathcal{M}\Delta B-\tfrac{2}{m}q\Delta\Phi} \label{eq:downstreamshockparallel}.
\end{equation}
For a reflected particle, the term under the root sign is negative. Here $\Delta B=B_{2}-B_{1}$, $q$ is the charge and $m$ is the mass of the particle and \mbox{$\mathcal{M}\equiv (m (1-\mu_{1}^{2})v_{1}^{2})/(2 B_1)$}. Thus, using the magnetic compression ratio $r_{B}=B_2/B_1$, any particle for which
\begin{equation}
u_{1}^{2}-2\mu vu_{1}+v^{2}-v^{2}(1-\mu^{2})r_{B}-2\tfrac{q}{m}\Delta\Phi \le 0 \label{eq:reflectioncondition}
\end{equation}
holds true is reflected at the shock. As the particle travels across the shock front, the parallel kinetic energy reaches zero and the electrostatic potential and the magnetic gradient cause the particle to be deflected in such a way that it is returned to the upstream. The details of reflection-threshold velocities are examined in more detail in Appendix \ref{appsec:analytical}, and speed thresholds for reflection, described in equations \eqref{eq:magnref1} and \eqref{eq:magnref2}, are shown in figure \ref{fig:thresholds}. 

\subsection{Particle transmission across the shock} \label{sec:parttransfer}

Assuming particle injection after transmission occurs through isotropisation (see section \ref{sec:isotropy} and equation \ref{eq:returnprob}), the primary parameter that defines the injection probability for a transmitted particle is the downstream plasma frame speed $v'$. 

\begin{figure}[!t]
\centering
\resizebox{\hsize}{!}{\includegraphics{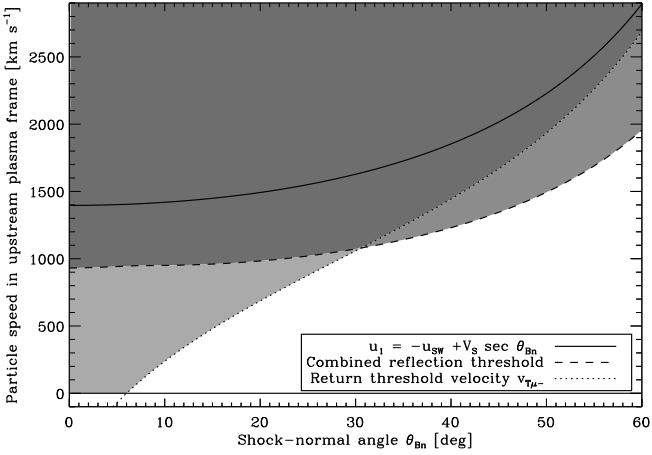}}%pics/speedthresholds2_vs1500
\caption{Injection threshold speeds for $V_\mathrm{s}=1500\,\mathrm{km\,s}^{-1}$: The classical threshold $u_1$ (solid curve) versus our reflection threshold (dashed) and the threshold for injection after downstream scatterings (dotted, $v > v_{\mathrm{T}\mu_-}$). Protons within the shaded areas have nonzero injection probability; the different shading represents the injection possibility through reflection, downstream scatterings, or both. The classical threshold overestimates the lowest injection speeds, especially at small shock-normal angles.}
\label{fig:returnthresholds}
\end{figure}

We find the particle downstream perpendicular velocity as
\begin{equation}
v'_\perp = v'\sqrt{1-\mu'^{2}} = v\sqrt{1-\mu^{2}}\sqrt{r_{B}}\label{eq:downstreamperp}
\end{equation}
and the particle downstream plasma frame parallel velocity as
\begin{equation}
v'_\parallel = v'\mu' = u_{2}-\sqrt{u_{1}^{2}-2\mu vu_{1}+v^{2}-v^{2}(1-\mu^{2})r_{B}-2\tfrac{q}{m}\Delta\Phi}.\label{eq:downstreamparallel}
\end{equation}
These can be used to solve the downstream plasma frame particle speed $v'$ and pitch-angle $\mu'$. The equations for $v'$ and $\mu'$ can be further analysed to study whether a particle transmitted to the downstream can return to the upstream. We discuss and derive analytical expressions for the threshold velocities for this return in Appendix \ref{appsec:analytical}. Speed thresholds for injection through reflection or downstream scatterings are shown in figure \ref{fig:returnthresholds} for a shock of velocity $V_\mathrm{s} = 1500\,\mathrm{km s}^{-1}$.

\section{Numerical flux analysis} \label{sec:numflux}

Although analytical methods can give information on threshold velocities for seed particle injection, they cannot be used directly to estimate the injection efficiency. To do this, we need to evaluate the proportion of the incoming particle flux, with a given upstream distribution function, that is returned to the upstream after its first encounter with the shock, that is, injected. We used a semi-analytical method, where the incident flux is found by integrating the flux-weighted upstream seed particle distribution as 
\begin{equation}
\int \frac{\mathrm{d}\mathcal{F}_{x_1=0}}{\mathrm{d}^3 v} \, d^3v.
\end{equation}
The distributions used in our model are described in equations \eqref{eq:fluxatsmallvelocities} and \eqref{eq:modifiedflux2}. The reflected flux is found by integrating over all cells where the requirements for reflection are fulfilled. The transmitted portion of the distribution function is conserved according to Liouville's theorem, with new downstream values $v'$ and $\mu'$ given by equations \eqref{eq:downstreamperp} and \eqref{eq:downstreamparallel}. The downstream flux is then found by integrating over the whole downstream distribution. We integrated the distribution with $v_{\parallel,1}$ and $v_{\parallel,2}$ spanning up to $10 u_1$ and $v_{\perp,1}$ and $v_{\perp,2}$ spanning up to $5 u_1$. Conservation of the shock-normal component of particle flux is confirmed.

\begin{figure*}[!t]
\centering
\includegraphics[width=\columnwidth]{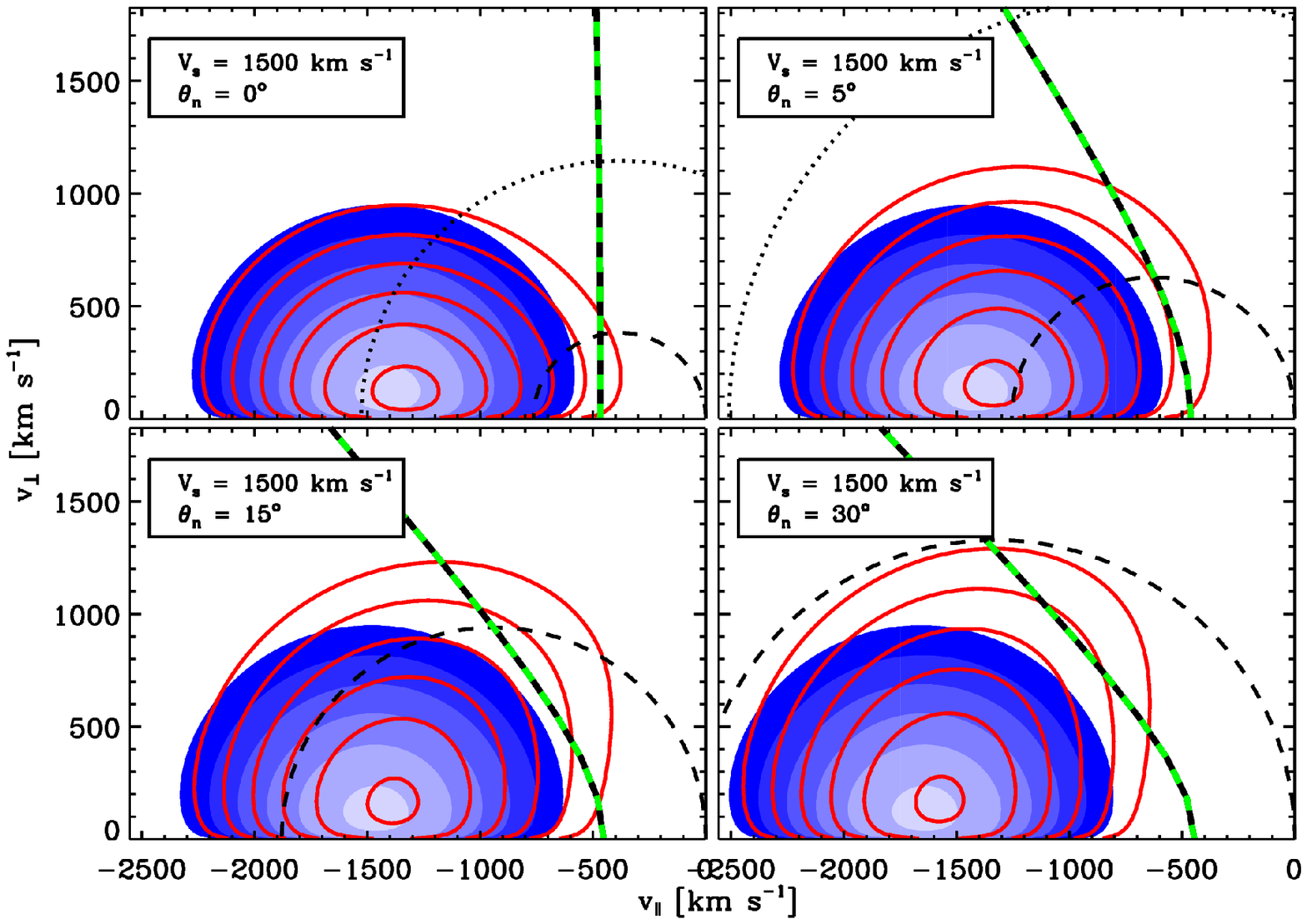}
\includegraphics[width=\columnwidth]{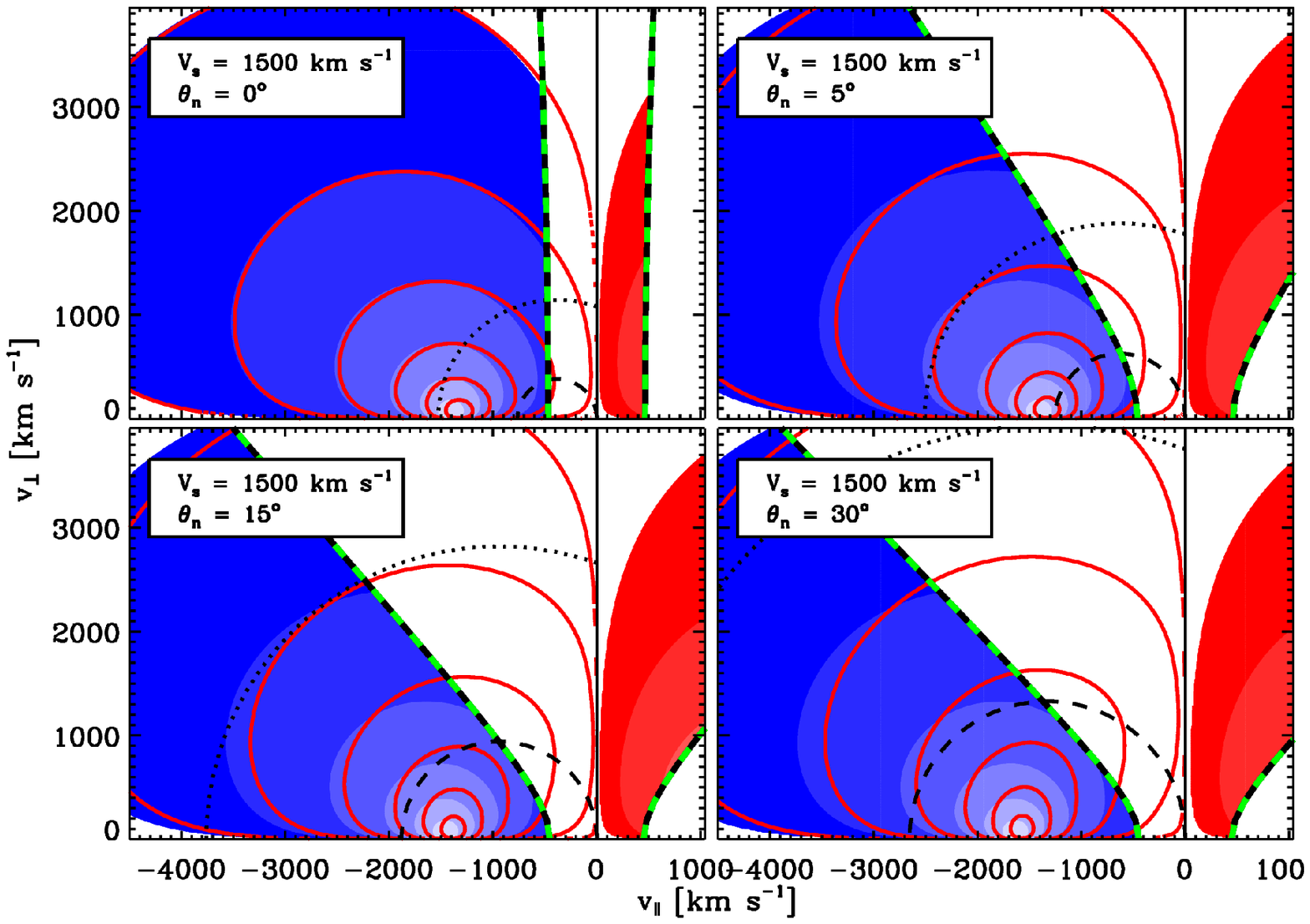}%pics/nofly_cartesian_k15_1500_raster,pics/nofly_cartesian_k2_1500_raster
\caption{Visualised transfer of quasi-thermal (left, $\kappa=15$) and non-thermal (right, $\kappa=2$) particle flux densities $\mathrm{d}\mathcal{F}/(\mathrm{d}v_\parallel \, \mathrm{d}v_\perp)$ across a shock of velocity $V_\mathrm{s}=1500\,\mathrm{km\,s}^{-1}$ with four different shock-normal angles. Blue contours depict the incident transmitting flux, red contours depict the reflected flux, and red solid contours the downstream transmitted flux at one magnitude intensity intervals. The black semicircles depict downstream speed thresholds for statistical return probabilities of 0\% (dashed) and 25\% (dotted), and the black-green dashed curve represents the reflection threshold.}
\label{fig:nofly_vs1500}%
\end{figure*}

\begin{figure*}[!t]
\centering
\includegraphics[width=\columnwidth]{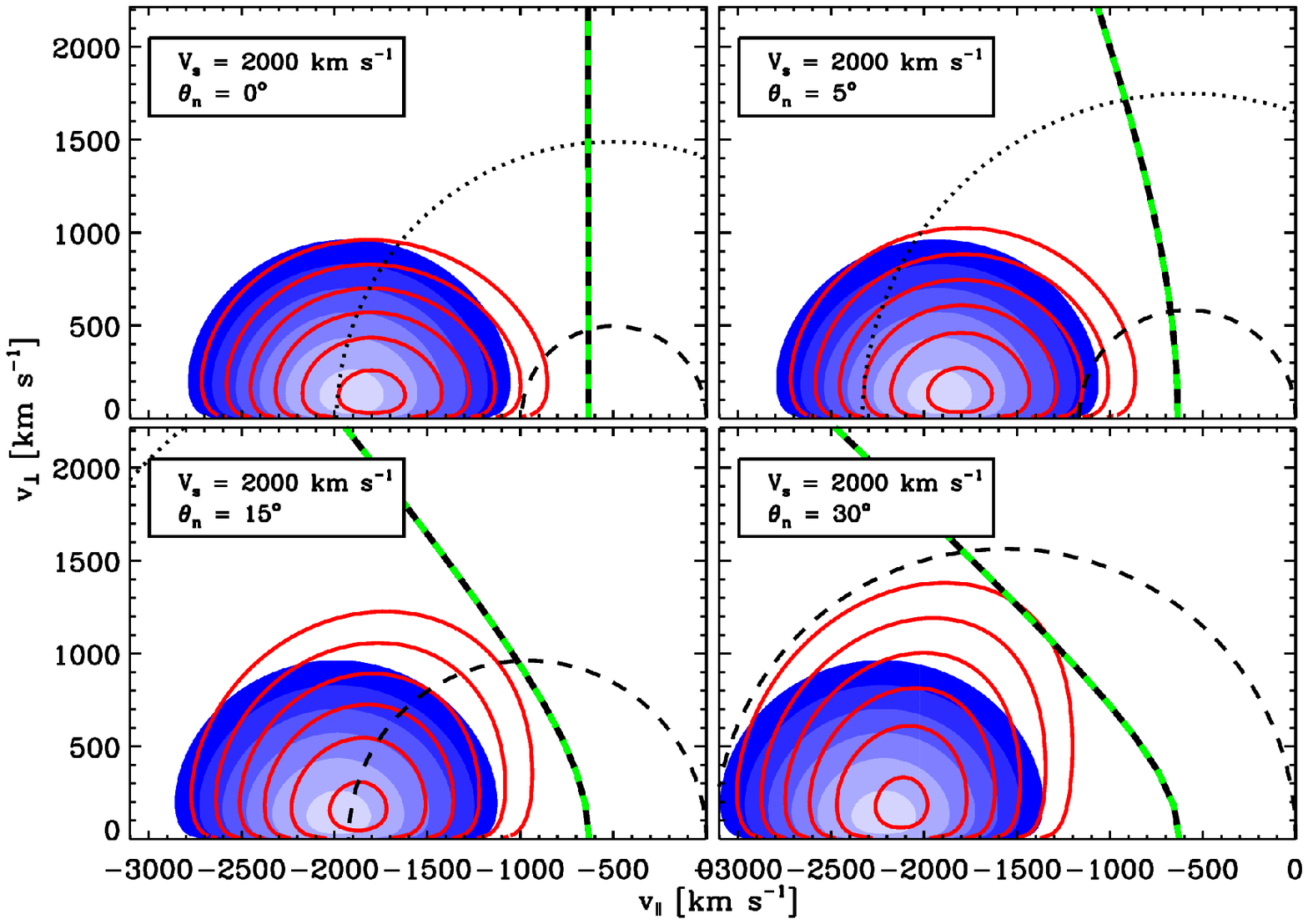}
\includegraphics[width=\columnwidth]{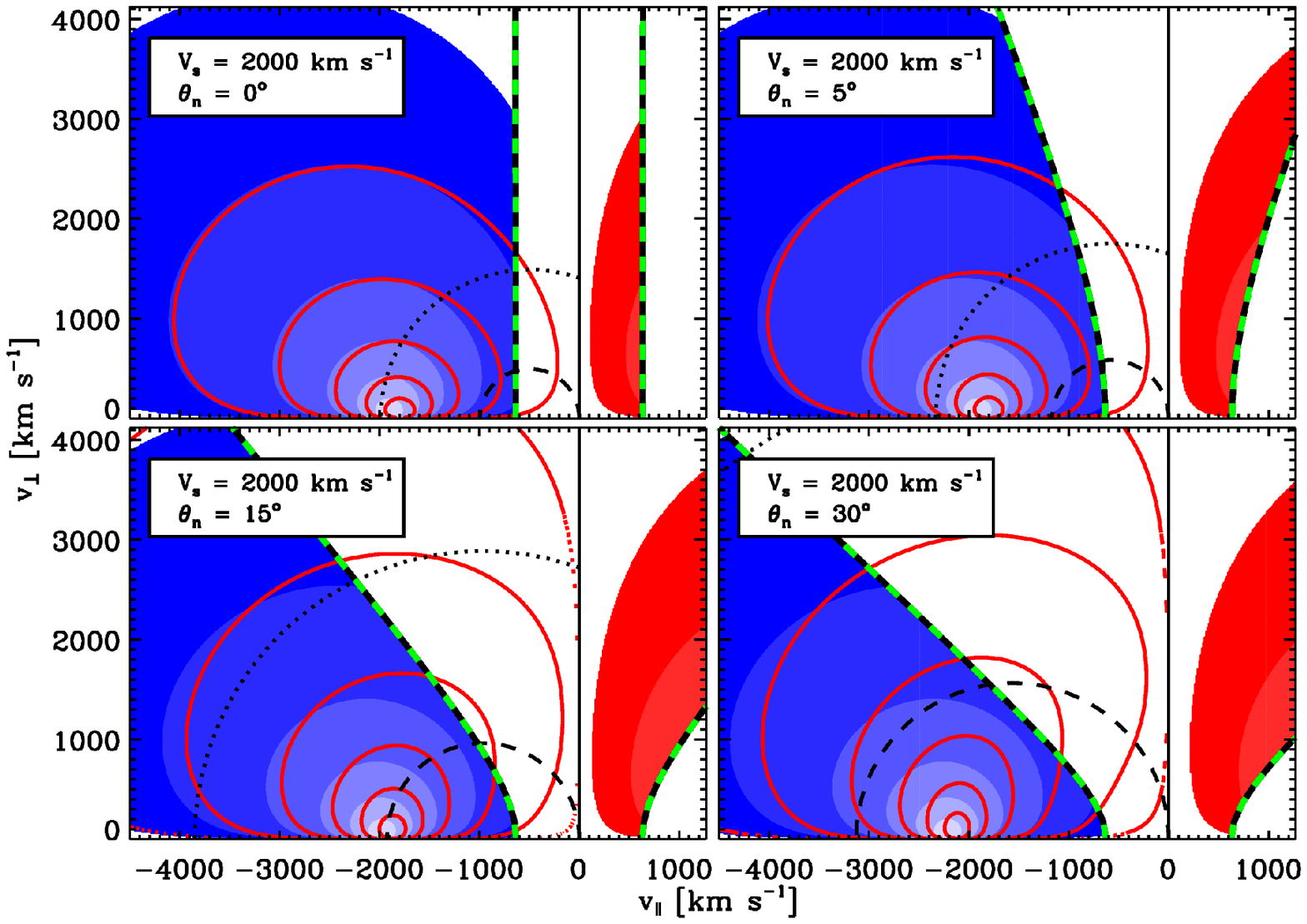}%pics/nofly_cartesian_k15_2000_raster,pics/nofly_cartesian_k2_2000_raster
\caption{Visualised transfer of quasi-thermal (left, $\kappa=15$) and non-thermal (right, $\kappa=2$) particle flux densities $\mathrm{d}\mathcal{F}/(\mathrm{d}v_\parallel \, \mathrm{d}v_\perp)$ across a shock of velocity $V_\mathrm{s}=2000\,\mathrm{km\,s}^{-1}$ with four different shock-normal angles. Blue contours depict the incident transmitting flux, red contours depict the reflected flux, and red solid contours the transmitted downstream flux at one magnitude intensity intervals. The black semicircles depict downstream speed thresholds for statistical return probabilities of 0\% (dashed) and 25\% (dotted), and the black-green dashed curve represents the reflection threshold.}
\label{fig:nofly_vs2000}%
\end{figure*}

In figures \ref{fig:nofly_vs1500} and \ref{fig:nofly_vs2000}, we present visualisations of quasi-thermal and non-thermal particle populations at two different shock-normal velocities and a selection of shock-normal angles. In the figures, velocities are measured in the frame of the shock. The shown quantity is the two-dimensional flux density $\mathrm{d}\mathcal{F}/(\mathrm{d}v_\parallel \, \mathrm{d}v_\perp)$, with contour boundaries at one-magnitude intervals. 

The transmitted portion of the incident flux is shown as blue filled contours. The threshold curve of reflection at the shock is shown as a black and green dashed line and the reflected flux is shown as red filled contours moving away from the shock in the upstream direction. The transmitted flux, modified by the shock passage as given by Eqs. \eqref{eq:downstreamperp} and \eqref{eq:downstreamparallel}, is shown as red contours at one-magnitude intervals. The downstream velocities of the transmitted particle flux can then be assessed for its injection probability according to equation \eqref{eq:returnprob}. The black dashed semicircle depicts the speed $u_2$ in the downstream medium, representing the lowest particle speed for a possible injection. A guideline representing the speed at which the statistical return probability for an isotropic population in an infinite downstream is 25 \% ($v' = 3u_2$) is shown as a black dotted semicircle.

\section{Monte Carlo simulations} \label{sec:montecarlo}

In this section, we describe Monte Carlo simulations that allow us to investigate the effect of downstream isotropisation and cross-field diffusion on the injection efficiency. A large number of particles were picked from the $\kappa$-distribution based on the flux conservation calculations presented in section \ref{sec:seedpopulations} and Appendix \ref{appsec:fluxweighting}. The upstream velocity and pitch-angle were randomised using equations \eqref{eq:kappa}, \eqref{eq:fluxweightmu1} and \eqref{eq:fluxweightmu2}. Owing to simulation run time constraints, the amount of simulated quasi-thermal particles was scaled down because the injection probability decreased with increasing shock-normal angle $\theta_{B\mathrm{n}}$.

As discussed in section \ref{sec:isotropy}, scatter-free transmission across the shock results in the downstream population being anisotropic. To study the effect of anisotropy in the downstream, we present four sets of simulations. In the first set, particles are assumed to achieve instant isotropy without scattering or propagation. In sets two through four, particles isotropise through propagation and scattering in the downstream. In set two, downstream cross-field diffusion is suppressed. In sets three and four it is active, with $a_\perp=0.1$. In set three, downstream cross-field diffusion can scatter the particle up to the vicinity of the shock, but not across it. In set four, the boundary condition at the shock allows particles with $\mu_2>0$ to propagate across the shock front to the upstream.

For simplicity, all our Monte Carlo simulations assumed a downstream of infinite extent. We propagated and scattered particles in the downstream within a region $x_\mathrm{n,2}\in(-\mathcal{B},0)$ where $\mathcal{B}$ is a boundary (parallel to the shock front) at which particles are considered to have isotropised (see also \citealp{1996ApJ...473.1029E}).

For set one (no downstream scattering), we set $\mathcal{B}=0$. For sets two through four, we used the diffusion equation $u_x \partial_x n = \partial_x \kappa_x \partial_x n$ to define the downstream return boundary at three times the diffusion distance $\kappa_x/u_x$ from the shock. The shock-normal diffusion coefficient $\kappa_x = \langle(\Delta x)^2\rangle/(2\Delta t)$ can be split into field-parallel and field-perpendicular components with
\begin{equation*}
\kappa_x = \kappa_\parallel \cos^2 \theta_{B\mathrm{n},2} +  \langle\kappa_\perp(\mu')\rangle \sin^2 \theta_{B\mathrm{n},2}.
\end{equation*}
The average downstream cross-field diffusion coefficient $\langle\kappa_\perp\rangle$ can be solved as
\begin{equation*}
\langle \kappa_\perp\rangle = \frac{1}{2} \int_{-1}^{+1} \kappa_\perp (\mu')\,\mathrm{d}\mu' =\frac{2}{3} a_\perp \kappa_\parallel.
\end{equation*}
Because the true pitch-angle distribution in the downstream is unknown, a cautious choice is instead to use the highest value. Using $\left.\kappa_\perp\right|_\mathrm{max}=a_\perp \kappa_\parallel$ and $\kappa_\parallel = v'\lambda_\parallel /3$, we found the position of the downstream return boundary to be
\begin{equation}
\mathcal{B}=\lambda_\parallel \left(\cos^2 \theta_{B\mathrm{n},2} +a_\perp \sin^2 \theta_{B\mathrm{n},2} \right) \frac{v'}{u_2 \cos \theta_{B\mathrm{n},2}}.
\end{equation}
We defined the downstream cross-field diffusion strengths as \mbox{$a_\perp=0$} (set two) and  \mbox{$a_\perp=0.1$} (sets three and four).

In the downstream, the field-parallel particle motion is
\begin{equation*}
\delta x_\parallel = (v'\mu' -u_2)\,\delta t.
\end{equation*}
The diffusive step across the mean field is solved using the stochastic differential equation method, as described in \cite{gardiner1985handbook}. Thus, the particle is propagated with
\begin{equation*}
\delta x_\perp = \mathcal{R}_\mathrm{n}\sqrt{2 \kappa_\perp(\mu')
 \delta t} = \mathcal{R}_\mathrm{n}\sqrt{(1-\mu'^{2})(2/3)a_\perp \lambda_\parallel v' \,\delta t},
\end{equation*}
where $\mathcal{R}_\mathrm{n}$ is a normal distributed random number with unit variance and zero mean. We additionally constrained the field-perpendicular motion to $\delta x_\perp \leq v_\perp' \delta t$, which implies that we cut off the tail of the normal distribution of $\mathcal{R}_\mathrm{n}$ at $\mathcal{R}_\mathrm{n}=\sqrt{(3 v'\delta t')/(2 a_\perp \lambda_\parallel)}$. Combining $\delta x_\parallel$ and $\delta x_\perp$, the motion normal to the shock front is
\begin{equation*}
\delta x_\mathrm{n} = \delta x_\parallel \cos \theta_{B\mathrm{n},2} + \delta x_\perp \sin \theta_{B\mathrm{n},2}.
\end{equation*}

Particles encountering the boundary $x_\mathrm{n} = -\mathcal{B}$ because of the field-parallel motion were considered isotropic, and the return probability $P_\mathrm{ret}$ from equation \eqref{eq:returnprob} was applied. The particle was then sent back towards the shock from the distance of \mbox{$x_\mathrm{n} = -\mathcal{B}$} with a randomised flux-weighted shock-bound pitch-angle. In sets $2\ldots4$, particles propagated and experienced small-angle scattering in the downstream and are possibly again convected to \mbox{$x_\mathrm{n} = -\mathcal{B}$}. Each time this occurred, equation \eqref{eq:returnprob} was applied, which resulted in the cumulative return probability $P_\mathrm{c} = \prod P_{\mathrm{ret},i} $. Downstream propagation continued until the particle either returned to the shock or the cumulative return probability dropped below $10^{-6}$, at which point the particle was considered to be not injected. If the particle returned to the shock, the successful particle was injected with the assigned probability $P_\mathrm{c}$.

Because the rejection of uninjected particles may result in significantly reduced statistics for the injected flux, we used the minimum-variance unbiased estimator (see, e.g., \citealp{lehmann1998theory}) to improve the statistical accuracy of the injection probability. The method is described in Appendix~\ref{appsec:estimator}.

For downstream cross-field diffusion, we had to specify the terminating conditions for the trajectories that tried to cross boundaries at $x_\mathrm{n} = -\mathcal{B}$ or  $x_\mathrm{n} = 0$. We note that for cross-field diffusion across the downstream return boundary, there should be an equivalent, symmetric cross-field-diffusion flux backwards if the particle density is constant. We maintained this symmetry by cancelling out all steps $\delta x_\perp$ that would result in a particle crossing the downstream return boundary.

We assumed that the perpendicular mean free path in the upstream region is zero, which means that no particles can propagate into the upstream by perpendicular diffusion. Thus, we needed to specify a terminating condition for the trajectories that tried to enter the upstream region via cross-field diffusion. We modelled two different boundary conditions at the shock front. In set three, we prevented particle propagation into the upstream through cross-field diffusion by reflecting all perpendicular steps $\delta x_\perp$ at the shock front. In set four, we modified the boundary condition to allow cross-field jumps where the particle parallel motion is towards the upstream (i.e. where $\mu_2>0$) to inject the particle. The method of set four is intended to model the way in which cross-field diffusion brings the particle close to the shock, after which it is free to propagate along the field into the upstream.

\section{Results and discussion}

\subsection{Velocity thresholds and shock obliquity} \label{sec:velocityobliquity}

In figure \ref{fig:thresholds}, we show the analytical threshold velocity for reflection as a function of shock obliquity for a shock with $V_\mathrm{s}=1500\,\mathrm{km\,s}^{-1}$. As can be seen, reflection is only possible for particles with initial speeds higher than $\sim 930\,\mathrm{km\,s}^{-1}$, with the threshold increasing with shock-normal angle. For $V_\mathrm{s}=2000\,\mathrm{km\,s}^{-1}$, we found a similar behaviour, with the threshold starting at $\sim 1260\,\mathrm{km\,s}^{-1}$. Thus, through reflection, our fast shocks are capable of injecting only suprathermal particles.

The threshold for transmitted particle return to the upstream is not so straightforward, because for some shocks, particles of all velocities can be returned to upstream. In figure \ref{fig:returnthresholds}, we show that injection for transmitted particles of negligible upstream plasma frame speed ($v=0$) is possible with shock-normal angles $\theta_{B\mathrm{n}} \leq 6^\circ$ when $V_\mathrm{s}=1500\,\mathrm{km\,s}^{-1}$. For $V_\mathrm{s}=2000\,\mathrm{km\,s}^{-1}$ the requirement is $\theta_{B\mathrm{n}} \leq 14^\circ$. Because the bulk of the seed particles is thermal and thus relatively slow, these analytical results suggest a significant effect relating the efficiency of shock acceleration with the shock-normal angle $\theta_{B\mathrm{n}}$. For comparison, in addition to the lowest injection speed of transmitted particles, figure \ref{fig:returnthresholds} shows the classical injection speed threshold $u_1$ and speed thresholds for reflection. The classical model of $v>u_1$ overestimates the lowest injection speeds by several hundred $\mathrm{km\,s}^{-1}$.

The semi-analytical flux transmission visualisations presented in figures \ref{fig:nofly_vs1500} and \ref{fig:nofly_vs2000} confirm that an increasing shock-normal angle $\theta_{B\mathrm{n}}$ results in a decreasing injection probability for the core thermal population. In the parallel case ($\theta_{B\mathrm{n}}=0^\circ$), the core is found to be close to the 25 \% injection probability threshold. For $V_\mathrm{s} = 1500 \,\mathrm{km\,s}^{-1}$, the threshold of no injection for a shock-normal angle of $\theta_{B\mathrm{n}}=5^\circ$ approaches the thermal core, and a shock-normal angle of $\theta_{B\mathrm{n}}=15^\circ$ results in the highest three magnitudes of flux density falling within the zone of no return. For $V_\mathrm{s} = 2000 \,\mathrm{km\,s}^{-1}$, a similar effect is seen at slightly larger shock-normal angles. At $\theta_{B\mathrm{n}}=15^\circ$, the thermal core is intersected by the threshold of no return, and $\theta_{B\mathrm{n}}=30^\circ$ sees the highest six flux density magnitudes within the zone of no return.

\begin{figure*}[!t]
\centering
\includegraphics[width=\columnwidth]{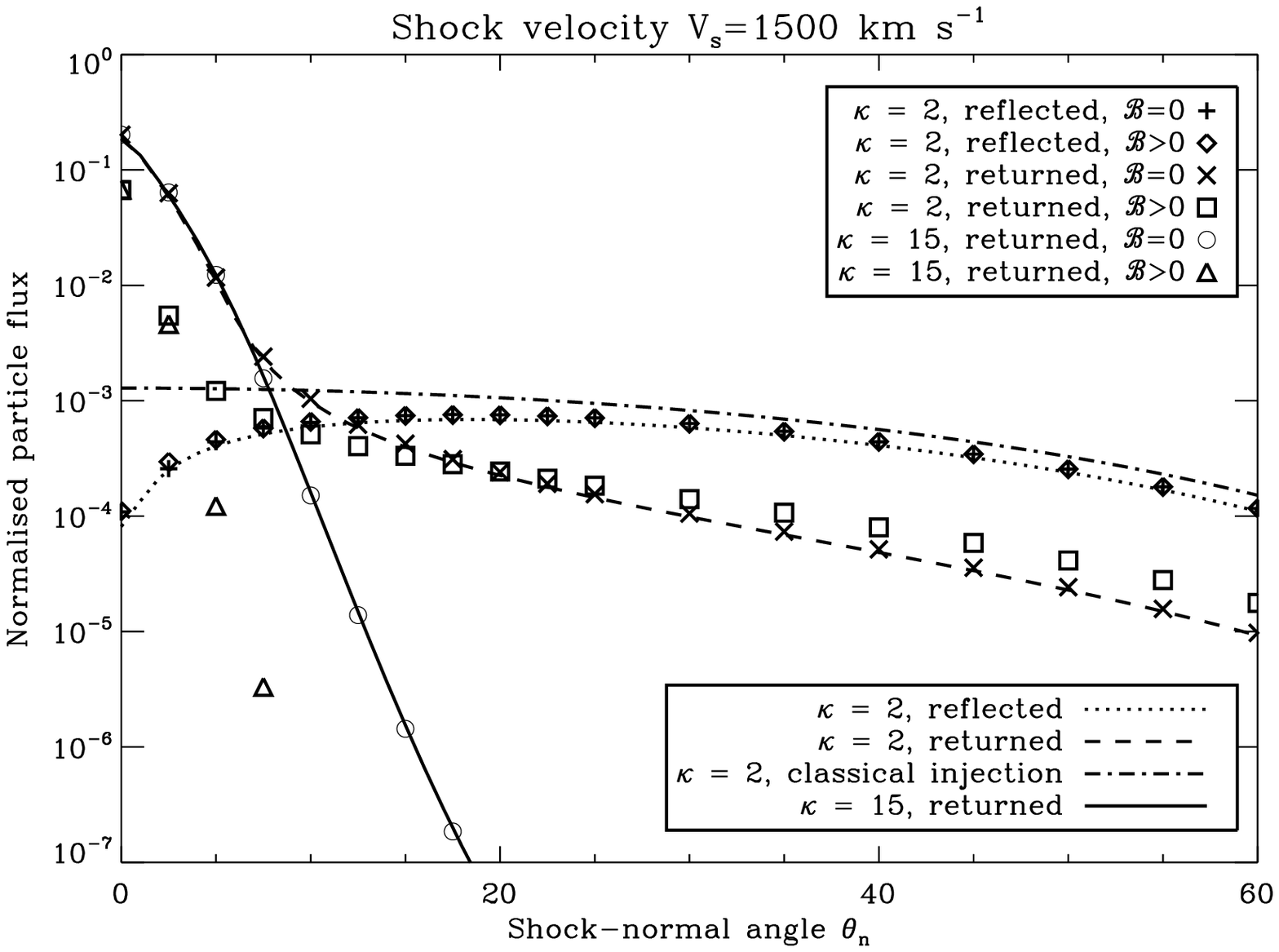}
\includegraphics[width=\columnwidth]{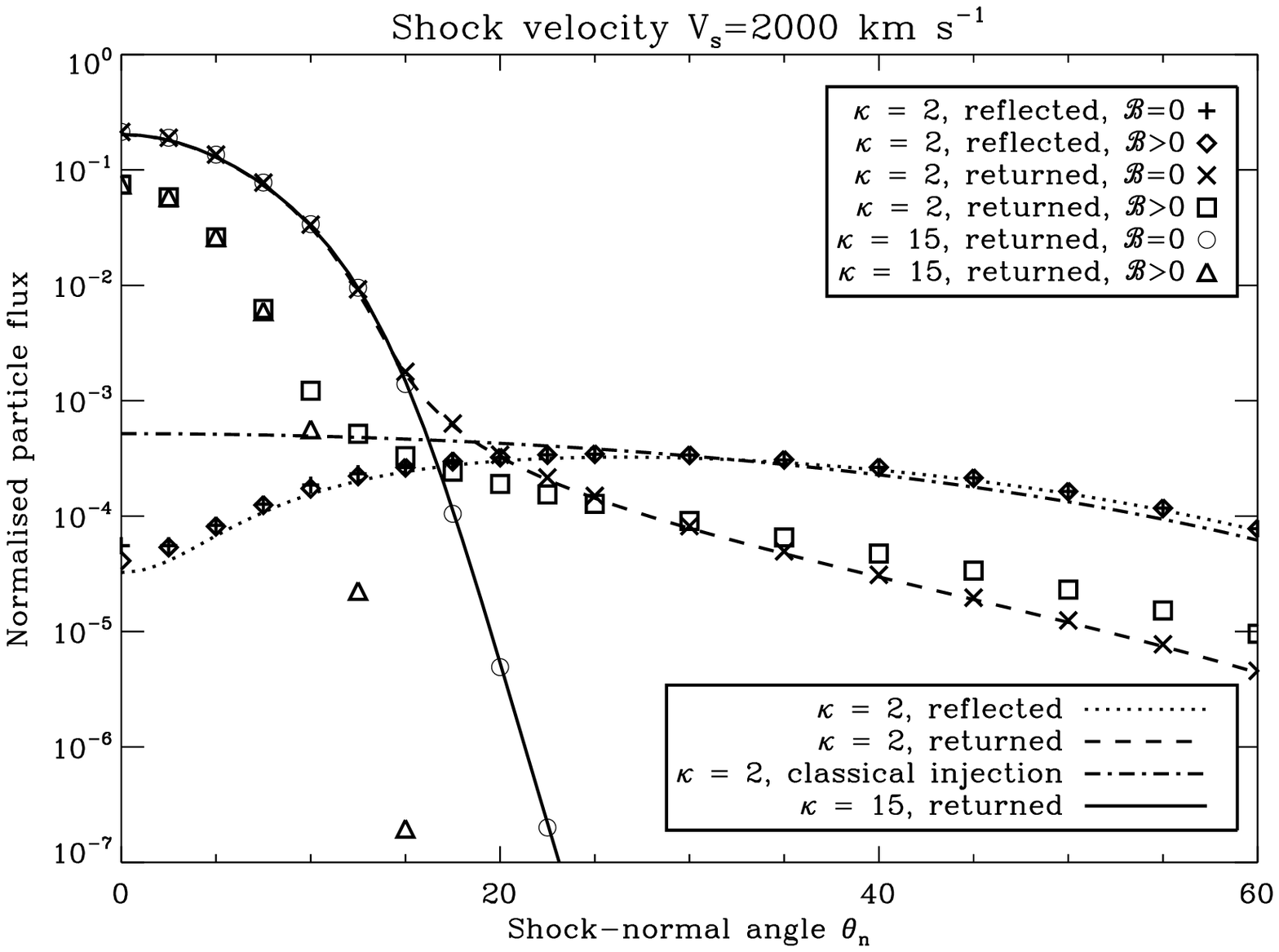}%pics/injection_flux14mono_comparisons_vs1500_csp,pics/injection_flux14mono_comparisons_vs2000_csp
\caption{Particle fluxes at the shock: Curves represent numerical integration results, data points are results of Monte Carlo simulations. $V_\mathrm{s} = 1500 \,\mathrm{km\,s}^{-1}$ (left) and $V_\mathrm{s} = 2000 \,\mathrm{km\,s}^{-1}$ (right). Simulations without downstream propagation of particles are designated with $\mathcal{B}=0$, whereas simulations where particles isotropise through downstream scatterings are designated with $\mathcal{B}>0$. The flux differences for the reflected particles in the Monte Carlo simulations ($\diamond,+$) are due to statistical uncertainties. For comparison, we show the integrated flux of all particles that are faster than the classical injection threshold $v>u_1$ as a dash-dotted line. For $\kappa=15$, reflected particle fluxes and the classical comparison are too low to be visible.}
\label{fig:fluxcomparisons}%
\end{figure*}

Our main results are shown in figure \ref{fig:fluxcomparisons}, where we compare the results of Monte Carlo simulations with semi-analytical results. We calculated injected fluxes for particles both reflected at the shock and injected through downstream scatterings. For the semi-analytical flux analysis, this was achieved by integrating the selected $\kappa$-distribution over the whole velocity space, accounting for reflection or return probability. The classical threshold result was found by integrating the population for all particles that fulfil $v>u_1$. When reflection dominates, this comparison is quite close to our results, but it should be noted that injection of all particles with $v>u_1$ is unrealistic, with the difference compensated for because our model is able to also reflect particles of upstream speeds $v<u_1$. For Monte Carlo simulations, the representative weight of all injected (reflected or returned) particle packages was compared with the total weight of all tested particles.

We found that Monte Carlo simulations without downstream propagation yield injection probabilities in line with semi-analytical results. As the shock-normal angle $\theta_{B\mathrm{n}}$ increases, a threshold effect begins to prevent the abundant, slow-core particle population from being injected. For $V_\mathrm{s} = 1500 \,\mathrm{km\,s}^{-1}$, an injection efficiency decrease of two orders of magnitude is seen as the shock-normal angle $\theta_{B\mathrm{n}}$ increases from $0$ to $7.5^{\circ}$. For $V_\mathrm{s} = 2000 \,\mathrm{km\,s}^{-1}$, a similar decrease is observed as $\theta_{B\mathrm{n}}$ increases from $0$ to $15^{\circ}$. We note that this effect is greater than the one reported by \cite{1995ApJ...453..873E}, spanning several orders of magnitude. However, our work does not retain a constant gas compression ratio, but keeps $V_\mathrm{s}$ constant, thus allowing the compression ratio to decrease as obliquity increases.

For the non-thermal distribution ($\kappa=2$), the reflected particle flux becomes higher than the flux of transmitted and subsequently injected particles when $\theta_{B\mathrm{n}} \geq 13^\circ$ ($V_\mathrm{s} = 1500 \,\mathrm{km\,s}^{-1}$) or when $\theta_{B\mathrm{n}} \geq 21^\circ$( $V_\mathrm{s} = 2000 \,\mathrm{km\,s}^{-1}$). For the quasi-thermal distribution ($\kappa=15$), the reflected particle flux is negligible and the injection probability continues to decrease as $\theta_{B\mathrm{n}}$ increases. Injection of a quasi-thermal distribution is found to be negligible at $\theta_{B\mathrm{n}} \geq 25^\circ$.

We also studied the case of shock-normal velocity $V_\mathrm{s} = 1000 \,\mathrm{km\,s}^{-1}$, which did not show a strong correlation between injection efficiency and shock-normal angle. With such a low shock-normal velocity, we found the injection of the thermal core to be improbable, regardless of the shock-normal angle $\theta_{B\mathrm{n}}$. This is mostly due to the small gas compression ratio for such a low Mach number shock.

\subsection{Isotropisation in the downstream} \label{sec:isotropisation}

In figure \ref{fig:fluxcomparisons}, we show how downstream isotropisation methods affect our Monte Carlo simulations. We show that at small (large) shock-normal angles, the assumption of instant downstream isotropy results in a higher (lower) injection probability than the one actually needed to allow the population to achieve isotropy through downstream scattering. This occurs because with velocities $v' > u_2$, the average parallel velocity in the downstream plasma frame $v'_\parallel$ for transmitted particles is negative (positive) at small (large) shock-normal angles. Additionally, the thermal population exclusion is seen to be activated at smaller shock-normal angles if isotropisation takes place through simulated scatterings. For example, at $V_\mathrm{s} = 1500 \,\mathrm{km\,s}^{-1}$ and $\theta_{B\mathrm{n}}=10^\circ$, accounting for downstream scattering decreases the returning flux of transmitted particles by well over three orders of magnitude. At $V_\mathrm{s} = 2000 \,\mathrm{km\,s}^{-1}$ and $\theta_{B\mathrm{n}}=10^\circ$, the decrease is approximately two orders of magnitude. At parallel ($\theta_{B\mathrm{n}}=0^\circ$) shocks, the returning flux was found to decrease by 33 \% ($V_\mathrm{s} = 1500 \,\mathrm{km\,s}^{-1}$) or by 35 \% ($V_\mathrm{s} = 2000 \,\mathrm{km\,s}^{-1}$) compared with the values found by assuming instant downstream isotropy.

Thus, it is evident that the effects of transmitted population anisotropy cannot be ignored when gauging the injection efficiency at a propagating coronal shock.

\subsection{Cross-field diffusion} \label{sec:res-crossfield}

\begin{figure}[!t]
\centering
\resizebox{\hsize}{!}{\includegraphics{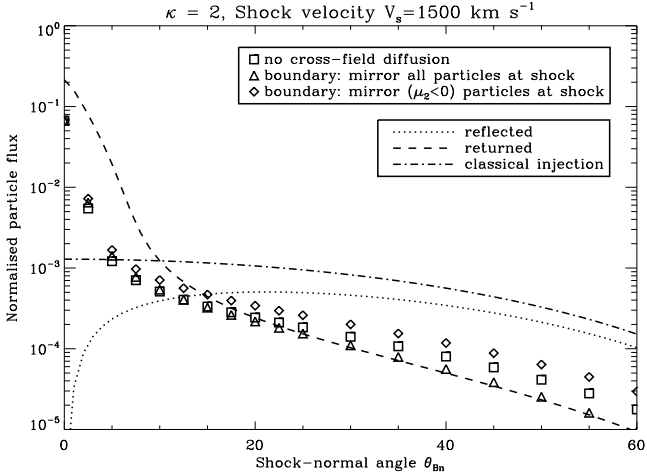}}%pics/injection_crossfield_k2_vs1500_csp
\caption{Effect of downstream cross-field diffusion strength $a_\perp=0.1$ on the injection efficiency for $V_\mathrm{s} = 1500 \,\mathrm{km\,s}^{-1}$, $\kappa=2$, and different shock-normal angles. Curves (from numerical integration results) and results from simulations without cross-field diffusion (squares) are the same as in figure \ref{fig:fluxcomparisons}. Results for the two types of downstream cross-field diffusion (sets three and four) are shown as triangles and diamonds.}
\label{fig:crossfield}
\end{figure}

In figure \ref{fig:crossfield}, we show comparisons between the simulation sets $2\ldots4$, which shows that accounting for the downstream cross-field diffusion affects the injection efficiency. At very low values of $\theta_{B\mathrm{n}}$, the cross-field diffusion in the downstream, acting as if it were boosting the isotropisation speed, increases the injection efficiency, but by a small margin. As the downstream shock-normal angle $\theta_{B\mathrm{n},2}$ increases, downstream cross-field diffusion begins to play a larger role, and the choice of boundary conditions at the shock becomes very important. Allowing downstream cross-field diffusion to take place, but preventing jumps of particles with $\mu_2>0$ to the upstream, acts like rapid isotropisation, and the injection efficiency approaches the $\mathcal{B}=0$ case. However, allowing downstream cross-field diffusion to inject particles with $\mu_2>0$ increases the injection by a factor of up to $\sim50\,\%$. At large shock-normal angles reflection remains the dominant injection mechanism.

\subsection{Cross-shock potential effect on injection}\label{sec:csp}

\begin{figure*}[!t]
\centering
\includegraphics[width=\columnwidth]{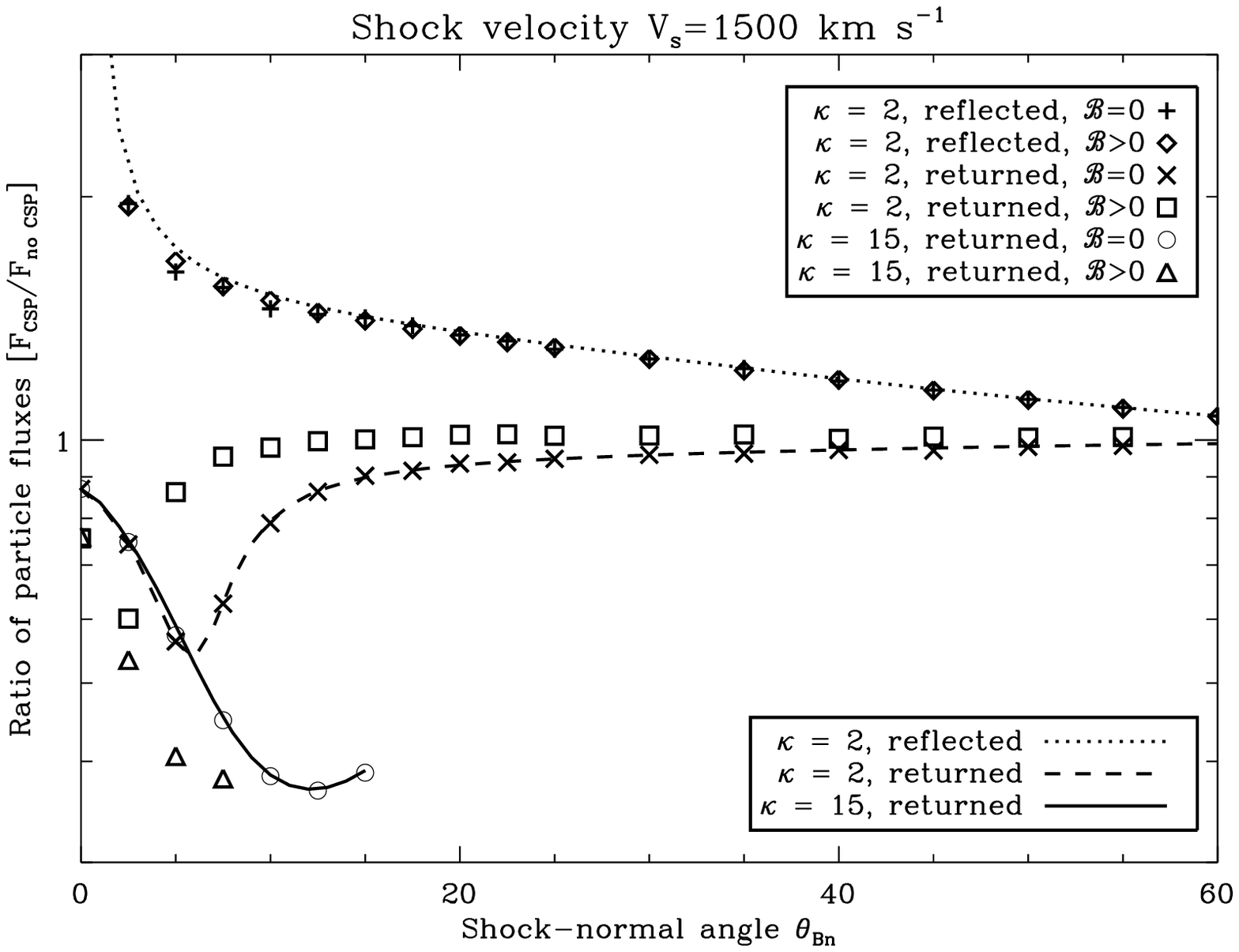}
\includegraphics[width=\columnwidth]{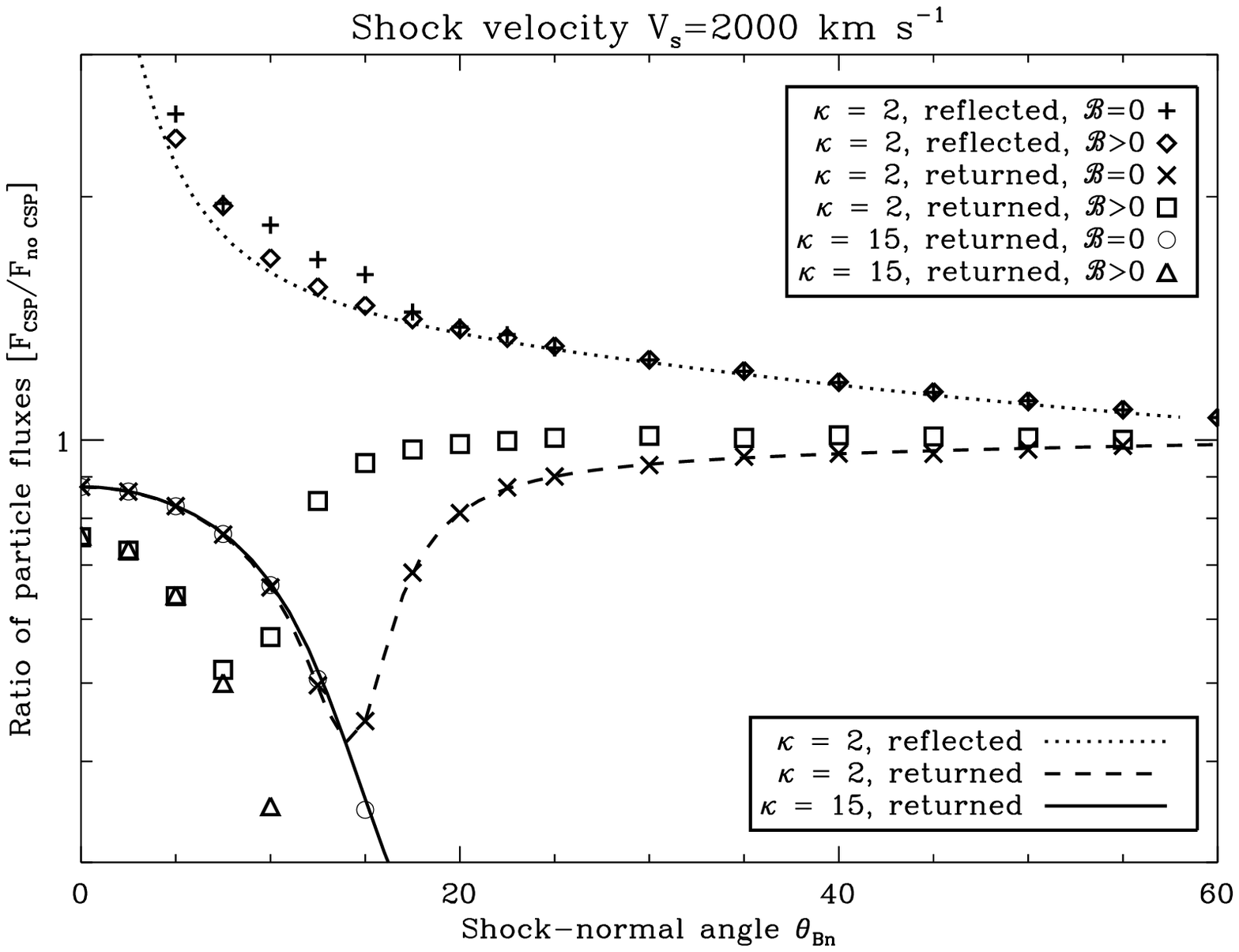}%pics/csp_ratio14mono_vs1500,pics/csp_ratio14mono_vs2000
\caption{Ratios of total injected particle fluxes with active ($F_\mathrm{CSP}$) or inactive ($F_\mathrm{no CSP}$) cross-shock potential. We show results for semi-analytical integration (curves) and Monte Carlo simulations (symbols). Presented shock-normal velocities are $V_\mathrm{s} = 1500 \,\mathrm{km\,s}^{-1}$ (left) and $V_\mathrm{s} = 2000 \,\mathrm{km\,s}^{-1}$ (right). Simulations without downstream propagation of particles are designated with $\mathcal{B}=0$, whereas simulations where particles isotropise through downstream scatterings are designated with $\mathcal{B}>0$. The flux differences for reflected particles in the Monte Carlo simulations ($\diamond,+$) are due to statistical uncertainties. For $\kappa=15$, reflected particle fluxes are too low to be visible.}
\label{fig:csp_ratio}%
\end{figure*}

In the presence of a cross-shock electric potential, particles are more likely to be reflected from the shock, and transmitted particles experience a decrease in parallel velocity. Thus, the expectation is for transmitted particles to be less likely to experience injection. Injected particles do, however, receive an additional speed boost when returning to the upstream. Injection efficiency ratios with active or inactive cross-shock potential are presented in figure \ref{fig:csp_ratio}. Results are shown only when the flux in question is not lower than $10^{-6}$ of the incoming flux. Reflection at small shock-normal angles in the absence of a cross-shock potential is very unlikely, which results in the large statistical uncertainty in Monte Carlo simulations.

Both the semi-analytical approach and Monte Carlo simulations confirm the increase in reflection probability and decrease in transmitted particle injection probability. As $\theta_{B\mathrm{n}}$ increases, and injection of reflected particles is possible only for the non-thermal tail of the $\kappa=2$ population, the effect of the cross-shock potential on the injection diminishes. Interestingly, accounting for downstream scattering causes the diminishing of thermal particle injection to occur at significantly smaller shock-normal angles than if assuming instant downstream isotropy. This is because the average parallel velocity in the downstream plasma frame $v'_\parallel$ for transmitted particles (with velocities $v' > u_2$) changes direction.

As the shock-normal angle $\theta_{B\mathrm{n}}$ increases, the magnetic reflection probability of particles increases, and the relative magnitude of the effect caused by the cross-shock potential decreases. For $V_\mathrm{s} = 1500 \,\mathrm{km\,s}^{-1}$, the increase of reflected particles at $\theta_{B\mathrm{n}}=5^\circ$ is approximately 65 \%. For $V_\mathrm{s} = 2000 \,\mathrm{km\,s}^{-1}$, the increase of reflected particles at $\theta_{B\mathrm{n}}=5^\circ$ is approximately 270 \%. Both cases see the increase drop to approximately 30 \% when $\theta_{B\mathrm{n}}$ approaches $25^\circ$.

As mentioned in section \ref{sec:velocityobliquity}, reflection overtakes downstream scattering as a source of injected particles when  $\theta_{B\mathrm{n}} \geq 14^\circ$ ($V_\mathrm{s} = 1500 \,\mathrm{km\,s}^{-1}$), or when $\theta_{B\mathrm{n}} \geq 21^\circ$ ($V_\mathrm{s} = 2000 \,\mathrm{km\,s}^{-1}$). Subsequently, at small (large) shock-normal angles, the net injection effect of a cross-shock potential is negative (positive).

%______________________________________________________________

\section{Conclusions}

We have presented analytical and numerical results that indicate that quasi-parallel coronal shocks are capable of injecting a significant portion of the cold core of thermal particle when the shock-normal velocity $V_\mathrm{s}$ is sufficiently high. Requiring injected particles to fulfil the classical injection threshold $u_1 = V_\mathrm{s} \sec \theta_{B\mathrm{n}} -u_\mathrm{sw}$ is unlikely to accurately describe particle injection.

With the assistance of Monte Carlo simulations, we showed the $\theta_{B\mathrm{n}}$-dependence of the injection efficiency to be highly sensitive to a proper assessment of transmitted particle flux anisotropy. Using an assumption of instant downstream isotropy can overestimate the injection probability by a factor of 3 to 3000. Allowing for downstream cross-field diffusion to take place can either increase or decrease the injection efficiency of particles, depending on chosen boundary conditions. However, cross-field diffusion in the downstream has a negligible effect at small shock-normal angles, where thermal particles can be injected. At large shock-normal angles, reflection remains the dominant injection method. Note, however, that our model assumed that cross-field diffusion is limited to the downstream region, indicating that this transport process can only bring the particle back to the shock, but not across it. If cross-field diffusion in the upstream region were added, its role in the injection process would most probably increase.

The DSA at coronal shocks is strongly dependent on the strength of self-generated turbulence, and thus, the amount of seed particles taking part in the acceleration process. The results presented in this paper suggest that fast quasi-parallel coronal shocks are capable of efficiently injecting the thermal core of the solar wind. A short, finite interval of parallel shock propagation could be efficient in boosting a significant portion of the thermal particle population to suprathermal energies, resulting in an injection hot-spot, after which they could be efficiently accelerated by either a quasi-parallel or an oblique shock front.

Additionally, we showed that a cross-shock potential has a significant effect in increasing the probability of reflection for incident particles and in decreasing the injection probability of thermal particles. For suprathermal particle populations, and by extension, oblique shocks, the cross-shock potential was shown to be of little significance.

%
%______________________________________________________________

\begin{acknowledgements}
        The calculations presented were performed using the Finnish Grid Infrastructure (FGI)
        project (Turku, Finland).
	The authors would like to thank %the IT Center for Science Ltd (CSC) for computational
%	services and 
        the Academy of Finland (AF) for financial support of projects 133723 and 259227. 
        TL acknowledges support from the UK Science and Technology Facilities Council (STFC)
        (grant ST/J001341/1). The research leading to these results has received funding from
        the European Union’s Seventh Framework Programme (FP7/2007-2013) under grant agreement
        No 262773 (SEPServer).
	The authors gratefully acknowledge the important comments and suggestions provided by the anonymous referee.
\end{acknowledgements}

%
%__________________________________________________________________

\bibliographystyle{aa} % style aa.bst
\bibliography{aa21348}

\appendix

\section{Flux weighting}\label{appsec:fluxweighting}

\subsection{Upstream conservation of flux} \label{sec:conservationofflux}

The flux of particles through a given cross-sectional area of the flux tube towards the shock is given as
\begin{equation}
\mathcal{F}\equiv\frac{\mathrm{d}N}{\mathrm{d}A\,\mathrm{d}t}=-\int\mathrm{d}^{3}v_{1}\mu_{1}v_{1}f=\int\mathrm{d}^{3}v\,(u_1-\mu v)\, f(x_1,v,\mu),
\end{equation}
which satisfies the steady state requirement of
\begin{equation*}
\frac{\partial\mathcal{F}}{\partial x_1}=0.
\end{equation*}
It is noteworthy that although the mean direction of particles is in the negative x-axis direction, we defined the total net flux across the shock as being positive. We considered the shock to be an absorbing boundary, that is, the distributions and fluxes represent particles that have yet to encounter the shock. The particle flux, which is constant regardless of position $x_1$, is easily evaluated far from the shock, where only the scattering process affects the angular distribution of the particles. The distribution is assumed to be isotropic in the local plasma frame. Thus, sufficiently far upstream, the flux is given by the form
\begin{equation}
\mathcal{F}=\int\mathrm{d}^{3}v\,(u_1-\mu v)\, f_{\infty}(v)=u_{1}n_{\infty},
\end{equation}
where $n_{\infty}$ is the particle density and $f_{\infty}$ the particle distribution at infinity. 

At velocities $v < u_1$, all particles travel towards the shock, and thus, information of the shock at $x_1=0$ cannot propagate into the upstream. The distribution remains isotropic in the local plasma frame, and the differential flux impacting the shock is simply
\begin{equation}
\frac{\mathrm{d}\mathcal{F}_{x_1=0}}{\mathrm{d}^{3}v}=(u_1-\mu v)\, f_{\infty}(v).\label{eq:fluxatsmallvelocities}
\end{equation}

Considering particles with speeds $v > u_1$, the picture becomes more complicated. Information of the shock can propagate against the flow of plasma, because $\mu v -u_1$ can be positive. This means that the angular distribution of particles that have not yet interacted with the shock becomes anisotropic, as the distribution $f(x_1=0,v,\mu) = 0$ for $\mu v >u_1$, due to the absorbing boundary at the shock. However, at a distance of $x_1=2\lambda$, for instance, we can still assume isotropy and take $f(2\lambda,v,\mu)\approx f_{\infty}(v)$, so that
\begin{equation}
\frac{\mathrm{d}\mathcal{F}_{x_1=2\lambda}}{\mathrm{d}^{3}v}\approx(u_{1}-\mu v)\, f_{\infty}(v)
\end{equation}
is valid for all velocities including $v>u_1$. In this flux, particles with $\mu v >u_1$ cause a negative contribution to the net flux, because they propagate outwards across the flux tube cross-section surface at $x_1=2\lambda$.

For the purpose of constructing a semi-analytical model of particle injection at high particle speeds, we simplified the anisotropies of the particle distribution at the shock. We assigned the modified differential flux of particles with $v>u_1$ at the shock as
\begin{equation}
\frac{\mathrm{d}\hat{\mathcal{F}}_{x_1=0}}{\mathrm{d}^{3}v}=\begin{cases}
(u_{1}-\mu v)\,\hat{f}(v), & \mu v<u_{1}\\
0, & \mu v>u_{1},\end{cases}\label{eq:modifiedflux1}
\end{equation}
where $\hat{f}(v)$ is a scaled distribution function yielding the correct total net flux, that is, $\hat{f}(v)=f_{\infty}(v)$ for $v<u_{1}$, but for $v>u_{1}$,
\begin{eqnarray*}
\int_{-1}^{u_{1}/v}\mathrm{d}\mu\,(u_{1}-\mu v)\,\hat{f}(v) & = & \int_{-1}^{+1}\mathrm{d}\mu\,(u_{1}-\mu v)\, f_{\infty}(v)\\
\Rightarrow\hat{f}(v) & = & \frac{4vu_{1}}{(v+u_{1})^{2}}f_{\infty}(v).
\end{eqnarray*}
Thus, for the particles that have not yet interacted with the shock, we find
\begin{equation}
\frac{\mathrm{d}\hat{\mathcal{F}}_{x_1=0}}{\mathrm{d}^{3}v}=\begin{cases}
\frac{4vu_{1}(u_{1}-\mu v)}{(v+u_{1})^{2}}f_{\infty}(v), & \mu v<u_{1}\\
0 & \mu v>u_{1},\end{cases}\label{eq:modifiedflux2}
\end{equation}
valid for particles of speed $v>u_{1}$. This accounts for the absorbing boundary at the shock whilst maintaining the conservation of total flux at a given particle speed $v$.

For Monte Carlo simulations, the same consideration of flux conservation must be made. The flux, that is, the total amount of particles encountered by the shock within time $\mathrm{d}t$, is formulated as
\begin{align}
\frac{\mathrm{d}N}{\mathrm{d}A\,\mathrm{d}t\,\mathrm{d}v} =
2\pi v^{2}\int_{-1}^{+1}\frac{\mathrm{d}\mathcal{F}_{x_1=\infty}}{\mathrm{d}^{3}v}\mathrm{d}\mu.
\end{align}
For particles of speeds $v < u_1$, the whole population is advected towards the shock, meaning that no information of the approaching shock can reach the particle distribution before impact. Thus, for these particles, the differential flux, extended to encompass all pitch-angles, can be written as
\begin{align}
\left. \frac{\mathrm{d}N}{\mathrm{d}A\,\mathrm{d}t\,\mathrm{d}v\,\mathrm{d}\mu} \right|_{v<u_{1}}
= 2\pi v^2 f_{\infty}(v) (u_1-\mu v).
\end{align}
Thus, the probability of a particle with speed $v$ exhibiting pitch-angle $\mu$ when impacting the shock is given as
\begin{align}
\left. P(\mu|v)\right|_{v<u_{1}} = \frac{u_1-\mu v}{2 u_1}, -1\,\leq\,\mu\,\leq\,+1.
\end{align}
This can be integrated to find the cumulative distribution function for a value of $\mu$ as
\begin{align}
 \int_{-1}^{\mu} \left. P(\mu'|v)\right|_{v<u_{1}} d\mu' = \frac{1+\mu}{2} + \frac{v}{4 u_1}(1-\mu^2).\label{eq:cumulative1}
\end{align}
From this, the Monte Carlo randomisation formula for $\mu$ can be solved as
\begin{equation}
\left. \mu\right|_{v<u_{1}} = \tfrac{u_1}{v} - \sqrt{(\tfrac{u_1}{v}-1)^2 +4\tfrac{u_1}{v}\mathcal{S} }, \label{eq:fluxweightmu1}
\end{equation}
where $\mu$ receives values from the range $-1 < \mu \leq +1$ and $\mathcal{S}$ is a uniformly distributed random number in the range $\mbox{[0,1)}$.

For particles with speeds $v > u_1$, information of the propagating shock can extend into the upstream, affecting the incident particle pitch-angle distribution. Thus, we initialised the particle distribution in the upstream of the shock at a distance of $x_1 = 2\lambda$, and allowed particles to convect towards the shock. This resulted in a realistic pitch-angle distribution at $x_1 = 0_{+}$, without having to resort to flux modification (equation \ref{eq:modifiedflux2}).

This pre-propagation is limited to the region $x_1 \in [0,2\lambda]$ with particles initialised isotropically at values $\mu < u_1/v$. As the total population is advected towards the shock, all particles that escape to $x_1 > 2\lambda$  will eventually return to the initialisation boundary of $x_1=2\lambda$, isotropised in the fluid frame. Thus, particles escaping to the upstream can be simply re-initialised at that position.

The distribution of pitch-angles $\mu$ for a given speed $v$, limiting the valid pitch-angle range to values $-1\,\leq\,\mu\,\le\,\tfrac{u_1}{v}$ and normalising the total probability to 1, is
\begin{align}
\left. P_\mathrm{pre}(\mu|v)\right|_{v>u_{1}} = \frac{2v\,(u_1-\mu v)}{(v+u_1)^2}, \,-1\,\leq\,\mu\,\leq\,\tfrac{u_1}{v}.\label{eq:fastmcpitchangle}
\end{align}
This will result in the shock-incident flux
\begin{equation}
\left. \frac{\mathrm{d}N}{\mathrm{d}A\,\mathrm{d}t\,\mathrm{d}v\,\mathrm{d}\mu}\right|_{v>u_{1}} = 2\pi v^2 f_{\infty}(v) \frac{4v u_1 (u_1 -\mu v)}{(v+ u_1)^2}, \quad \mu v<u_{1}.
\end{equation}
Using equation \eqref{eq:fastmcpitchangle}, the Monte Carlo randomisation formula for $\mu$ can be solved (similar to equation \ref{eq:cumulative1}) as
\begin{equation}
\left. \mu\right|_{v>u_{1}} = -\left(-\frac{u_{1}}{v}+(1+\frac{u_{1}}{v})\sqrt{\mathcal{S}}\right), \label{eq:fluxweightmu2}
\end{equation}
where $\mu$ receives values from the range $-1 < \mu \leq \tfrac{u_1}{v}$.

\section{Analytical injection thresholds} \label{appsec:analytical}

\subsection{Reflection threshold} \label{appsec:reflection}

In attempting to determine particle injection, we can solve certain seed particle speed thresholds. A particle is reflected (see Eq. \ref{eq:reflectioncondition}), and thus, injected, if 
\begin{equation}
2\mu vu_{1}-v^{2}+v^{2}(1-\mu^{2})r_{B}>u_{1}^{2}-2\tfrac{q}{m}\Delta\Phi. \label{eq:threshold1}
\end{equation}
The right-hand side (RHS) of the equation is constant. Through roots of derivatives of the left-hand side (LHS), we can find LHS maxima at $\mu=\tfrac{u_{1}}{r_{B}v},$ if $v>\tfrac{u_{1}}{r_{B}}$, or $\mu=1$, if $v\leq \tfrac{u_{1}}{r_{B}}$. Thus, if $v>\tfrac{u_{1}}{r_{B}}$, the LHS maximum is given as
\begin{equation}
2\frac{u_{1}^{2}}{r_{B}}-v^{2}+(v^{2}-\frac{u_{1}^{2}}{r_{B}^{2}})r_{B}=v^{2}(r_{B}-1)+\frac{u_{1}^{2}}{r_{B}}.
\end{equation}
This results in no possibility of reflection, if $u_{1}/r_{B} < v < v_{\mathrm{R}1}$, where
\begin{equation}
v_{\mathrm{R}1} = \sqrt{\frac{u_{1}^{2}}{r_{B}}-\frac{2q\Delta\Phi}{(r_{B}-1)m}}. \label{eq:magnref1}
\end{equation}
If $v\leq \tfrac{u_{1}}{r_{B}}$, the LHS has a maximum value of $2vu_{1}-v^{2}$. This results in no possibility of reflection, if $v < v_{\mathrm{R}2}$, where
\begin{equation}
v_{\mathrm{R}2} = u_{1}-\sqrt{2\tfrac{q}{m}\Delta\Phi}. \label{eq:magnref2}
\end{equation}

The LHS minimum is found at $\mu=-1$. With these shock and solar wind parameters, there exists no valid speed $v$ for which this value of the LHS would be positive. Under these circumstances, no seed particle speed $v$ results in certain reflection, regardless of pitch-angle $\mu$. 

\subsection{Threshold for return from the downstream} \label{appsec:return}

To split the transmitted particle population into portions with either possible or impossible injection, we examined equations \eqref{eq:downstreamperp} and \eqref{eq:downstreamparallel} to find
\begin{align}
v'^{2} &= v'^{2}_\perp+v'^{2}_\parallel = v^{2}\left(1-\mu^{2}\right)r_{B} \nonumber \\
 +& \left[u_{2}-\sqrt{u_{1}^{2}-2\mu vu_{1}+v^{2}-v^{2}(1-\mu^{2})r_{B}-2\tfrac{q}{m}\Delta\Phi}\right]^{2}. \label{eq:downstreamvelocity}
\end{align}
We found the maxima of downstream speed $v'$ for a given $v$, because this can result in an injection velocity threshold. Maxima for $v'$ can be found at $\mu=-1$ and $\mu=+1$ or by solving the roots of the derivative of $v'(\mu)$. Assigning
\begin{equation}
\mathcal{D}(\mu) = u_{1}^{2}-2\mu vu_{1}+v^{2}-v^{2}(1-\mu^{2})r_{B}-2\tfrac{q}{m}\Delta\Phi
\end{equation}
and finding the first derivative for $v'^2$ gives
\begin{eqnarray}
\frac{\partial v'^{2}}{\partial\mu} & = & \frac{\partial}{\partial\mu}\left[u_{2}-\sqrt{\mathcal{D}(\mu)}\right]^{2}-2\mu v^{2}r_{B}\\
 & = & \frac{2(u_{1}-\mu vr_{B})vu_{2}}{\sqrt{\mathcal{D}(\mu)}}-2vu_{1}. \label{eq:middleextrema}
\end{eqnarray}
Solving the roots provides an extremum, if $\mu\in(-1,+1)$ and if
\begin{align}
&(u_{1}-\mu vr_{B})^{2}u_{2}^{2} = u_{1}^{2} \,\mathcal{D}(\mu) \\
&\Rightarrow \mu = \frac{u_{1}}{vr_{B}}\pm \nonumber \\
&\sqrt{\frac{u_{1}^{2}}{v^{2}r_{B}^{2}}-\frac{u_{1}^{2}(r_{B}^{2}-r^{2})+v^{2}r^{2}(r_{B}-1)+2r^{2}\tfrac{q}{m}\Delta\Phi}{v^{2}r_{B}(r_{B}^{3}-r^{2})}}.
\end{align}

If $v<\tfrac{u_{1}}{r_{B}}$ holds true, only the solution with the minus sign is of interest. These, in addition to the possible extrema at $\mu=-1$ and $\mu=+1$, result in several possible threshold velocities.

\subsubsection{Extrema at $\mu=-1$ and $\mu=+1$} \label{appsec:extrema1}

The possible downstream velocity extrema at $\mu=-1$ and $\mu=+1$ can be solved by refining equation \eqref{eq:downstreamvelocity}. Formulating this as the threshold for no injection results in
\begin{equation}
v'^{2}=\left[u_{2}-\sqrt{(u_{1}\mp v)^{2}-2\tfrac{q}{m}\Delta\Phi}\right]^{2} < u_{2}^{2},
\end{equation}
where the term inside the square root is positive for all transmitted particles. Thus, a transmitted particle can return to the upstream only if
\begin{eqnarray}
& \sqrt{(u_{1}\mp v)^{2}-2\tfrac{q}{m}\Delta\Phi} - u_{2} > u_{2} \\
\Rightarrow & (u_{1}\mp v)^{2} > 2\tfrac{q}{m}\Delta\Phi + 4 u_2^2.
\end{eqnarray}
For particles with $v > u_1$, injection is always possible. For particles with $v < u_1$, we found the additional requirements of 
\begin{eqnarray}
v > v_{\mathrm{T}\mu_-} = \sqrt{ 2\tfrac{q}{m}\Delta\Phi + 4 u_2^2 } - u_{1} & (\mu = -1) \label{eq:vgtsqrt} \\
v < v_{\mathrm{T}\mu_+} = u_{1} - \sqrt{ 2\tfrac{q}{m}\Delta\Phi + 4 u_2^2 } = -v_{\mathrm{T}\mu_-} & (\mu = +1). \label{eq:vltsqrt}
\end{eqnarray}
For our parameters, only $v_{\mathrm{T}\mu_-}$ provides a valid condition for injection.

\subsubsection{Extrema at $\mu \neq \pm 1$} \label{appsec:extrema2}

Solving the thresholds speeds for the one or two extrema given in equation \eqref{eq:middleextrema} in an analytical fashion does not result in easily applicable equations. However, solving the velocity thresholds for these extrema in a numerical fashion revealed valid extrema only at shock-normal angles $\theta_{B\mathrm{n}} \leq 4^{\circ}$. At these shock-normal angles, the existing extrema at $\mu = \pm 1$ already allow theoretical return of particles regardless of their speed. Thus, these extrema do not affect the found particle return speed thresholds. However, in a more general case, they cannot be ignored.

\section{Statistical handling of uninjected particles} \label{appsec:estimator}

In Section~\ref{sec:montecarlo}, we presented a method for estimating the capability of a shock to inject particles using Monte Carlo simulations. In our method, we propagated the particle in the downstream until it was either injected into the upstream, or the cumulative probability of return at the downstream boundary fell below $10^{-6}$ and it was removed from the simulation as an uninjected particle. For some shock parameters, however, this method will result in very low statistics for the injected particles. To improve the statistical accuracy, we used the fact that successes and failures -- injections and non-injections -- are distributed according to a negative binomial distribution. When the numbers of successes, $\mathcal{R}$, and failures, $\mathcal{K}$, are known, the probability for success $\mathcal{P}$ can be evaluated using the minimum-variance unbiased estimator (see, e.g., \citealp{lehmann1998theory}), which gives
\begin{equation}
\mathcal{P} = \frac{\mathcal{R}-1}{\mathcal{R}+\mathcal{K}-1}.
\end{equation}
It should be noted, however, that the actual probability of injection for a particle is
\begin{equation}
P_\mathrm{inj} = \mathcal{P} P_\mathrm{s},
\end{equation}
where $P_\mathrm{s}$ is the injection probability associated with the last encountered success. 

To evaluate the unbiased estimator, we randomised particles in groups. For each particle within the group, the newly randomised values of $v$ and $\mu$ were used to test the particle for reflection, as explained in section \ref{sec:reflection}. Reflected particles were considered successes and have $P_\mathrm{s}=1$. Non-reflected particles are transmitted to the downstream, with $v'$ and $\mu'$ calculated according to equations \eqref{eq:downstreamperp} and \eqref{eq:downstreamparallel}. They were then followed in the downstream, as described in Section~\ref{sec:montecarlo}, until they were either injected, incrementing $\mathcal{R}$, or considered uninjected, incrementing $\mathcal{K}$. This was continued until $\mathcal{R}=5$. The last particle of the group was then injected with the probability $P_\mathrm{inj}$.

As a precaution against excessively low success probabilities, the group size was limited to $(\mathcal{R}+\mathcal{K})=10^{6}$. If this limit was reached and at least two successes were encountered, the values of $\mathcal{R}_i$ and $\mathcal{K}_i$ associated with the last encountered success were used to calculate $\mathcal{P}$. If the tests resulted in only 0 or 1 success, the group was considered to result in no injection.

With these methods, the injected weight of the Monte Carlo particle was found to be $W_\mathrm{inj} = W_\mathrm{seed} P_\mathrm{inj}$, where $W_\mathrm{seed}$ is the representative weight of upstream seed particles assigned to this group, and was found based on the plasma density. The total injected weight was then used to calculate the particle flux.

\end{document}